%% file: art2.tex
\pdfoutput=1

\RequirePackage[l2tabu, orthodox]{nag}

\documentclass[11pt,
final,
onecolumn,
]{article}

\usepackage[english]{babel}

\usepackage[T1]{fontenc} % Choose font encoding for output
\usepackage[utf8]{inputenc} % Choose font encoding for input
\usepackage{microtype}
\usepackage{lmodern} % Latin Modern font
\usepackage{csquotes}

\usepackage{amsmath, amsfonts, amssymb} % Load an extensive set of mathematical symbols
\let\origlangle\langle % Store original LaTeX \langle under a different name before loading mathabx
\let\origrangle\rangle % Store original LaTeX \rangle under a different name before loading mathabx
\usepackage[matha,mathx]{mathabx} % Obtain extensible accents like \widebar and a host of (modified) mathematical symbols. Note that the appearance of \langle also changes.
\let\langle\origlangle % Recover original LaTeX \langle
\let\rangle\origrangle % Recover original LaTeX \rangle

\usepackage[normalem]{ulem} % Allow for striking through of text
\usepackage{color}

\usepackage[totalwidth=480pt, totalheight=680pt]{geometry}

\usepackage[affil-it]{authblk}

\usepackage{graphicx}

\usepackage{booktabs}

\newcommand{\todo}[1]{{\bf \color{red} Todo: #1}} % Command to mark all text that still has to be edited
\renewcommand{\todo}[1]{}

\newcommand{\replace}[2]{{\bf \color{blue} \sout{#1}#2}} % Marks deletions and insertions
\renewcommand{\replace}[2]{{\bf \color{blue} #2}} % Marks new text, not showing deleted text
\renewcommand{\replace}[2]{#2} % Final version

\newcommand{\tr}[1]{\operatorname{tr}(#1)}  % Trace of a matrix, in which an operator name is used that will be typeset straight-up and in roman font
\renewcommand{\det}[1]{\operatorname{det}(#1)}  % Determinant of a matrix, in which an operator name is used that will be typeset straight-up and in roman font
\newcommand*\dif{\mathop{}\! \, \mathrm{d}} % Straight-up differential. May be replaced by \dif from the commath package

\newcommand{\units}[1]{ \mathrm{\ #1} }

\newcommand{\eg}{e.g.}
\newcommand{\ie}{i.e.}
\newcommand{\etal}{et al.}
\newcommand{\cf}{cf.}
\newcommand{\dg}{$^\dagger$}

\newcommand{\filter}[1]{ \widebar{#1} }
\newcommand{\filterlength}{ \delta }
\newcommand{\testfilter}[1]{ \tilde{#1} }
\newcommand{\testfilterlength}{ \testfilter{ \delta } }

\newcommand{\LESu}{ v }
\newcommand{\LESp}{ q }
\newcommand{\LESG}{ G }
\newcommand{\LESS}{ S }
\newcommand{\LESOmega}{ \Omega }

\newcommand{\avg}[1]{\langle #1 \rangle}

\newcommand{\taumod}{ \tau^{ \mathrm{ mod } } } % When in a subscript, use _{ \taumod } instead of _\taumod
\newcommand{\taumodsub}[1]{ \taumod_{ \mathrm{ #1 } } } % Similar comment
\newcommand{\taumoddev}{ \tau^{ \mathrm{ mod,dev } } } % ...
\newcommand{\taumoddevsub}[1]{ \tau^{ \mathrm{ mod,dev } }_{ \mathrm{ #1 }} }
\newcommand{\tauhatmod}{ \hat{ \tau }^{ \mathrm{ mod } } }
\newcommand{\kmod}{ k^{ \mathrm{ mod } } }

\newcommand{\Ptau}{ P_{ \tau } }
\newcommand{\Qtau}{ Q_{ \tau } }
\newcommand{\Rtau}{ R_{ \tau } }
\newcommand{\Dtau}{ D_{ \tau } }
\newcommand{\Ptaumod}{ P_{ \taumod } }
\newcommand{\Qtaumod}{ Q_{ \taumod } }
\newcommand{\Rtaumod}{ R_{ \taumod } }
\newcommand{\Dtaumod}{ D_{ \tau }^{ \mathrm{mod} } }
\newcommand{\Qtaumoddev}{ Q_{ \taumoddev } }
\newcommand{\Rtaumoddev}{ R_{ \taumoddev } }
\newcommand{\PGGT}{ P_{ G G^T } }
\newcommand{\QGGT}{ Q_{ G G^T } }
\newcommand{\RGGT}{ R_{ G G^T } }
\newcommand{\QS}{ Q_{ \LESS } }
\newcommand{\RS}{ R_{ \LESS } }

\newcommand{\Dmodsub}[1]{ D^{ \mathrm{ mod } }_{ \mathrm{ #1 } } }

\newcommand{\nue}{ \nu_{ \mathrm{ e } } } % Eddy viscosity
\newcommand{\nuesup}[1]{ \nue^{ \mathrm{ #1 } } } % Eddy viscosity with superscript to indicate a specific model eddy viscosity
\newcommand{\Csub}[1]{ C_{ \mathrm{ #1 } } } % Model constant

\newcommand{\ReNr}{Re} % Bulk Reynolds number
\newcommand{\Lref}{L} % Bulk characteristic length scale
\newcommand{\uref}{u_{\mathrm{ref}}} % Bulk characteristic velocity scale
\newcommand{\Retau}{Re_{\tau}} % Friction Reynolds number
\newcommand{\utau}{u_{\tau}} % Friction velocity
\newcommand{\channelHalfWidth}{d} % Channel half-width
\usepackage[
    backend=bibtex8,
    sortcites=true,
    sorting=nyt,
    sortlocale=en_US,
    style=numeric,
    maxbibnames=99, 
]{biblatex}
\DeclareNameAlias{default}{last-first}
\ifpdf
\usepackage[
    bookmarks=true,
    bookmarksnumbered=true,
    bookmarksopen=true,
    colorlinks=true,
    allcolors=blue,
    pdfborder={0 0 0}
]{hyperref}
\hypersetup{pdftitle={Physical consistency of subgrid-scale models for large-eddy simulation of incompressible turbulent flows},pdfauthor={Maurits H. Silvis, Ronald A. Remmerswaal, and Roel Verstappen}}
\fi

\usepackage[capitalize, sort&compress]{cleveref} %sort&compress

\title{\vspace{-0.6\baselineskip} Physical consistency of subgrid-scale models for large-eddy simulation of incompressible turbulent flows}
\author{Maurits H. Silvis\thanks{Email address: \href{mailto:m.h.silvis@rug.nl}{m.h.silvis@rug.nl}} }
\author{Ronald A. Remmerswaal}
\author{Roel Verstappen}
\affil{Johann Bernoulli Institute for Mathematics and Computer Science, University of Groningen, Nijenborgh 9, 9747 AG Groningen, The Netherlands\vspace{-0.8\baselineskip}}
\begin{document}

\maketitle

\input{art2text}

% % BibLaTeX style bibliography % %
\begingroup
\setlength\bibitemsep{0pt}  
\printbibliography
\endgroup

\end{document}

%% file: art2text.tex
% !TeX root = art2.tex

\vspace{-10pt}
\noindent The following article appeared in \textit{Phys.~Fluids} 29, 015105 (2017) and may be found at \url{http://dx.doi.org/10.1063/1.4974093}.
This article may be downloaded for personal use only. Any other use requires prior permission of the author and AIP Publishing.

\paragraph{Abstract} We study the construction of subgrid-scale models for large-eddy simulation of incompressible turbulent flows.
In particular, we aim to consolidate a systematic approach of constructing subgrid-scale models, based on the idea that it is desirable that subgrid-scale models are consistent with the 
mathematical and physical
properties of the Navier-Stokes equations and the turbulent stresses.
To that end, we first discuss in detail the symmetries of the Navier-Stokes equations, and the near-wall scaling behavior, realizability and dissipation properties of the turbulent stresses.
We furthermore summarize the requirements that subgrid-scale models have to satisfy in order to preserve these important mathematical and physical properties.
In this fashion, a framework of model constraints arises that we apply to analyze the behavior of a number of existing subgrid-scale models that are based on the local velocity gradient.
We show that these subgrid-scale models do not satisfy all the desired properties, after which we explain that this is partly due to incompatibilities between model constraints and limitations of velocity-gradient-based subgrid-scale models.
However, we also reason that the current framework shows that there is room for improvement in the properties and, hence, the behavior of existing subgrid-scale models.
We furthermore show how compatible model constraints can be combined to construct new subgrid-scale models that have desirable properties built into them.
We provide a few examples of such new models, of which a new model of eddy viscosity type, that is based on the vortex stretching magnitude, is successfully tested in large-eddy simulations of decaying homogeneous isotropic turbulence and turbulent plane-channel flow.

\section{Introduction}
\label{sec:intro}

Most practical turbulent flows cannot be computed directly from the Navier-Stokes equations, because not enough resolution is available to resolve all relevant scales of motion.
We therefore turn to large-eddy simulation (LES) to predict the large-scale behavior of incompressible turbulent flows.
In large-eddy simulation, the large scales of motion\replace{s}{} in a flow are explicitly computed, whereas effects of small-scale motions 
have to be
modeled.
The question is\replace{}{,} how to model these effects\replace{.}{?}
Several answers to this question can be found in the literature.
For example,
since the advent of computational fluid dynamics
many 
so-called
subgrid-scale models have been proposed and successfully applied to the simulation of a wide range of turbulent flows
(see,
\eg,
the encyclopedic work of Sagaut~\cite{sagaut06}).
Given the variety of models proposed in the literature, the question remains, however, what defines a well-designed subgrid-scale model\replace{.}{?}
Some authors have therefore taken a systematic approach of finding constraints for the construction of subgrid-scale models~\cite{speziale85,vremanetal94,oberlack97,vreman04,verstappen11,verstappen16} (also 
refer to
the extensive review by Ghosal~\cite{ghosal99}).
Most of these constraints are based on the idea that it is desirable that subgrid-scale models are consistent with important mathematical and physical properties of the Navier-Stokes equations and the turbulent stresses.
In the current work we aim to consolidate this systematic approach and provide a framework
for the assessment of existing and the creation of new subgrid-scale models for large-eddy simulation.

Constraints on the properties of subgrid-scale models come in several forms.
For example, 
it is well known that 
the Navier-Stokes equations
are 
% form 
invariant under certain
transformations, 
such as
instantaneous rotations of the coordinate system and the Galilean transformation~\cite{pope11}.
Such transformations, also referred to as symmetries, play an important physical role because they make sure that the description of fluids is the same in all inertial frames of reference. 
Furthermore, they relate to conservation and scaling laws~\cite{razafindralandyetal07}.
To ensure physical consistency,
one could therefore argue that
it is desirable that 
subgrid-scale models preserve the symmetries of the Navier-Stokes equations.
Speziale~\cite{speziale85} 
was the first to emphasize 
the importance of Galilean invariance of subgrid-scale models for large-eddy simulation. 
Later, Oberlack~\cite{oberlack97} formulated requirements to make subgrid-scale models compatible with all the symmetries of the Navier-Stokes equations.
An example of a class of models that was designed to preserve the symmetries of the Navier-Stokes equations can be found in the work of \replace{Razafindralandy, Hamdouni and Oberlack.}{Razafindralandy \etal}~\cite{razafindralandyetal07}.

One could furthermore argue that
it is desirable that subgrid-scale models share some basic properties with the 
true
turbulent stresses, 
such as the observed near-wall scaling behavior~\cite{chapmankuhn86},
certain 
dissipation properties
and realizability~\cite{vremanetal94}.
Examples of subgrid-scale models 
that exhibit the same near-wall scaling behavior as the turbulent stresses are given by the WALE model of Nicoud and Ducros~\cite{nicoudducros99}, the $\sigma$ model of \replace{Nicoud, Baya Toda, Cabrit, Bose and Lee}{Nicoud \etal}~\cite{nicoudetal11} and the S3PQR models of \replace{Trias, Folch, Gorobets and Oliva.}{Trias \etal}~\cite{triasetal15}.
The dissipation behavior of the turbulent stresses was studied by Vreman~\cite{vreman04}, who proposed a model that has a vanishing subgrid dissipation whenever the true turbulent stresses are not causing energy transfer to subgrid scales.
The QR model~\cite{verstappenetal10,verstappen11,verstappenetal14} 
and the recently developed anisotropic minimum-dissipation (AMD) model of \replace{Rozema, Bae, Moin and Verstappen}{Rozema \etal}~\cite{rozemaetal15}
were designed to exhibit a particular dissipation behavior
that leads to scale separation between large and small scales of motion.

The property of realizability of the turbulent stresses
pertains to subgrid-scale models that, unlike the eddy viscosity models mentioned so far, include a model for the
generalized subgrid-scale kinetic energy.
Examples of 
realizable models
are the gradient model~\cite{leonard74,clarketal79}
and the explicit algebraic subgrid-scale stress model (EASSM) of \replace{Marstorp, Brethouwer, Grundestam and Johansson.}{Marstorp \etal}~\cite{marstorpetal09}.
A 
feature of interest 
of these models
is that they
contain terms that are nonlinear in the local velocity gradient.
As a consequence they can describe other-than-dissipative
processes,
allowing us to go
beyond the (mostly) dissipative description of turbulent flows that is provided by eddy viscosity models.
For other studies of subgrid-scale models that are nonlinear in the local velocity gradient, refer to, for instance, 
Lund and Novikov~\cite{lundnovikov92}, Kosovi\'{c}~\cite{kosovic97}, Wang and Bergstrom~\cite{wangbergstrom05}, and Wendling and Oberlack~\cite{wendlingoberlack07}. 
For an extensive review of the use of nonlinear models 
in the context of the Reynolds-averaged Navier-Stokes (RANS) equations,
see Gatski and Jongen~\cite{gatskijongen00}.
The reader that seeks detailed background information about nonlinear constitutive equations and their role in describing fluid flows in general is referred to the book by Deville and Gatski~\cite{devillegatski12}.

In the current paper we provide a detailed discussion of the aforementioned mathematical and physical properties of the Navier-Stokes equations and the turbulent stresses,
and we focus on the constraints that subgrid-scale models have to satisfy in order to preserve these properties.
We furthermore apply the framework that so arises
to perform a systematic analysis of the behavior of a number of existing subgrid-scale models.
Also, we illustrate how new subgrid-scale models can be designed 
that have desired properties built into them.
A few examples of such new models are provided, 
of which a model of eddy viscosity type is tested in numerical simulations of decaying homogeneous isotropic turbulence and a turbulent channel flow.
We note here that,
apart from the near-wall scaling requirements,
all the 
model constraints
that are discussed
in this paper
arise from 
analytical, deterministic considerations.
Also the assessment of existing subgrid-scale models is based on their analytical properties\replace{
and not on their behavior in a numerical simulation}{}.
For information about conditions on the 
statistical 
properties
of subgrid-scale models, 
refer to Langford and Moser~\cite{langfordmoser99}.
Also see the work by Meneveau and Marusic~\cite{meneveaumarusic13}, and \replace{Stevens, Meneveau and Wilczek.}{Stevens \etal}~\cite{stevensetal14}.

The outline of this paper is as follows.
In \cref{sec:LES} we introduce the Navier-Stokes equations and the equations underlying large-eddy simulation.
We furthermore discuss a class of subgrid-scale models
based on the local velocity gradient.
\cref{sec:modconstraints} is dedicated to the discussion of
mathematical and physical
properties of the Navier-Stokes equations and the turbulent stresses, as well as the resulting requirements for the form of subgrid-scale models.
An analysis of the properties of some existing subgrid-scale models is performed in \cref{sec:modanalysis}.
After that, in \cref{sec:modexamples}, we provide examples of new models that arise from the discussed requirements, 
along with numerical tests of a new eddy viscosity model.
Finally, \cref{sec:concldisc} consists of a summary of the current work and an outlook.

\section{Large-eddy simulation}
\label{sec:LES}

To facilitate the discussion of the properties of the Navier-Stokes equations and the turbulent stresses, 
and their consequences for subgrid-scale modeling,
we will first introduce the equations that underlie
large-eddy simulation.
In this section
we also introduce a general class of subgrid-scale models 
based on the local velocity gradient.

\subsection{The basic equations of large-eddy simulation}
\label{sec:LESeqs}

The behavior of constant-density Newtonian fluids at constant temperature is governed by the incompressible Navier-Stokes equations~\cite{pope11},
\begin{equation}
\label{eq:ns}
\frac{ \partial u_i }{ \partial t } 
+ \frac{ \partial }{ \partial x_j }( u_i u_j )
= -\frac{ 1 }{ \rho } \frac{ \partial p }{ \partial x_i } 
+ \nu \frac{ \partial^2 u_i }{ \partial x_j \partial x_j }, 
\qquad
\frac{ \partial u_i }{ \partial x_i } = 0.
\end{equation}
Here, $u_i$ represent the $x_i$-component of the velocity field
of the flow
and $p$ indicates the pressure. 
The 
density and kinematic viscosity are denoted by $\rho$ and $\nu$\replace{}{, respectively}.
Einstein's summation convention is assumed for repeated indices.

As remarked before, most practical turbulent flows cannot be computed directly from the Navier-Stokes equations, \cref{eq:ns},
because generally not enough resolution is available to resolve all relevant scales of motion.
We therefore turn to large-eddy simulation for the prediction of the large-scale behavior of turbulent flows.
In large-eddy simulation,
the distinction between large and small scales of motion is usually made by a filtering or coarse-graining operation.
This operation will be indicated by an overbar in what follows and
is assumed to commute with differentiation.

The evolution of incompressible large-scale velocity fields can 
formally
be 
described by the filtered incompressible Navier-Stokes equations~\cite{sagaut06},
\begin{equation}
\label{eq:filtns}
\frac{ \partial \filter{ u }_i }{ \partial t } 
+ \frac{ \partial }{ \partial x_j }( \filter{ u }_i \filter{ u }_j )
= -\frac{ 1 }{ \rho } \frac{ \partial \filter{ p } }{ \partial x_i } 
+ \nu \frac{ \partial^2 \filter{ u }_i }{ \partial x_j \partial x_j } 
- \frac{ \partial }{ \partial x_j } \tau_{ij}, 
\qquad
\frac{ \partial \filter{ u }_i }{ \partial x_i } = 0.
\end{equation}
The turbulent, or subfilter-scale, stresses, $\tau_{ij} = \filter{ u_i u_j } - \filter{ u }_i \filter{ u }_j$, represent the interactions between large and small scales of motion.
As they are not solely expressed in terms of the large-scale velocity field they have to be modeled.

In large-eddy simulation,
one looks for models for the turbulent stresses, $\taumod_{ij}( \LESu )$, such that the set of equations given by
\begin{equation}
\label{eq:les}
\frac{ \partial v_i }{ \partial t } 
+ \frac{ \partial }{ \partial x_j }( v_i v_j )
= -\frac{ 1 }{ \rho } \frac{ \partial q }{ \partial x_i } 
+ \nu \frac{ \partial^2 v_i }{ \partial x_j \partial x_j } 
- \frac{ \partial }{ \partial x_j } \taumod_{ij}( \LESu ), 
\qquad
\frac{ \partial v_i }{ \partial x_i } = 0,
\end{equation}
provides
accurate approximations for the filtered velocity, $v_i \approx \filter{ u }_i$, and pressure, $q \approx \filter{ p }$.
We will refer to \cref{eq:les} as the 
(basic) equations of large-eddy simulation.
We have purposely dropped the overbars in these equations because 
we will focus on large-eddy simulation without explicit filtering.
For this reason, we will refer to $\taumod_{ij}$ as a subgrid-scale stress model, or subgrid-scale model.
The subsequent discussion does, however, 
carry over to the case of subfilter-scale stress modeling for explicitly filtered large-eddy simulations, 
provided that the chosen filter satisfies the requirements discussed by Oberlack~\cite{oberlack97} and Razafindralandy
\etal~\cite{razafindralandyetal07}.

In practice, the large-eddy simulation equations, 
\cref{eq:les}, are solved numerically.
This step, which involves discretization
and is closely tied to the modeling process~\cite{verstappenetal14},
is not examined in detail in the current work.
Rather, we will focus on the analytical properties of the Navier-Stokes equations and the turbulent stresses, and discuss the constraints that subgrid-scale models have to satisfy to be consistent with these properties.
Before we continue that discussion, however, let us first introduce 
subgrid-scale models that are based on the local velocity gradient.

\subsection{Subgrid-scale models based on the local velocity gradient}
\label{sec:LESmodels}

As mentioned before, many different subgrid-scale models have been developed for large-eddy simulation~\cite{sagaut06}.
In the current work, 
after discussing the properties of the Navier-Stokes equations and the turbulent stresses, 
and focusing on the resulting model constraints,
we will investigate
subgrid-scale models that 
depend locally (\ie, without solving additional transport equations) on the
velocity gradient,
\begin{equation}
\label{eq:velgrad}
\LESG_{ij}( \LESu ) = \frac{ \partial \LESu_i }{ \partial x_j }.
\end{equation}
These subgrid-scale models can
be expressed in terms of the rate-of-strain and rate-of-rotation tensors,
\begin{equation}
\label{eq:SandW}
\LESS_{ij}( \LESu ) = \frac{ 1 }{ 2 }\left( \frac{ \partial \LESu_i }{ \partial x_j } + \frac{ \partial \LESu_j }{ \partial x_i } \right), 
\qquad\qquad 
\LESOmega_{ij}( \LESu ) = \frac{ 1 }{ 2 }\left( \frac{ \partial \LESu_i }{ \partial x_j } - \frac{ \partial \LESu_j }{ \partial x_i } \right).
\end{equation}
For brevity, we will write 
$\LESG_{ij} = \LESG_{ij}( \LESu )$,
$\LESS_{ij} = \LESS_{ij}( \LESu )$ and $\LESOmega_{ij} = \LESOmega_{ij}( \LESu )$ in what follows.
Where convenient we will furthermore employ matrix notation, dropping all indices.

The commonly used class of eddy viscosity models arises when it is assumed that small-scale turbulent motions effectively cause diffusion of the larger scales.
These models can be expressed as a linear constitutive relation between the deviatoric part of the subgrid-scale stresses,
\begin{equation}
\label{eq:taumoddev}
\taumoddev = \taumod - \frac{1}{3} \tr{ \taumod } I,
\end{equation}
and the rate-of-strain tensor, \ie,
\begin{equation}
\label{eq:modeddy}
\taumoddevsub{ e } = - 2 \nue S.
\end{equation}
The definition of the eddy viscosity, $\nue$, is discussed below.

Turbulence is described as an essentially dissipative process by eddy viscosity models.
To allow for the description of
other-than-dissipative
processes,
we will 
further
consider subgrid-scale models 
that contain tensor terms
that are nonlinear in the local velocity gradient.
A general class of models of that type
is given
by
\begin{equation}
\label{eq:modnonlin}
\taumod = \sum_{i = 0}^{10} \alpha_i T_i,
\end{equation}
where the tensors $T_i$ depend in the following way on the rate-of-strain and rate-of-rotation tensors~\cite{spencerrivlin58,spencerrivlin62,pope75,lundnovikov92}.
\begin{align}
\label{eq:tensors}
\begin{alignedat}{3}
T_0 & = I,           \qquad\qquad & T_4 & = \LESS\LESOmega - \LESOmega\LESS,                       \qquad\qquad & T_8 & = \LESS\LESOmega\LESS^2 - \LESS^2\LESOmega\LESS, \\
T_1 & = \LESS,       \qquad\qquad & T_5 & = \LESS^2\LESOmega - \LESOmega\LESS^2,                   \qquad\qquad & T_9 & = \LESS^2\LESOmega^2 + \LESOmega^2\LESS^2, \\
T_2 & = \LESS^2,     \qquad\qquad & T_6 & = \LESS\LESOmega^2 + \LESOmega^2\LESS,                   \qquad\qquad & T_{10} & = \LESOmega\LESS^2\LESOmega^2 - \LESOmega^2\LESS^2\LESOmega.\\
T_3 & = \LESOmega^2, \qquad\qquad & T_7 & = \LESOmega\LESS\LESOmega^2 - \LESOmega^2\LESS\LESOmega, \qquad\qquad & ~ & ~
\end{alignedat}
\end{align}
Here, $I$ represents the identity tensor. 
The model coefficients, $\alpha_i$, and also the eddy viscosity of \cref{eq:modeddy}, $\nue$, are generally defined as follows (no summation is implied over indices in brackets),
\begin{equation}
\label{eq:modcoefficients}
\alpha_i = c_{(i)} \filterlength^2 f_{(i)}(I_1, I_2, \ldots, I_5).
\end{equation}
That is, each of the model coefficients, $\alpha_i$, is
taken to be a product of three factors:
a dimensionless constant, $c_i$; 
a (squared) length scale, that is commonly associated with the subgrid characteristic (or filter) length scale of large-eddy simulation, $\filterlength$;
and a 
function $f_i$ \replace{}{with units of inverse time} that depends on the local velocity gradient through
the combined invariants of the rate-of-strain and rate-of-rotation tensors~\cite{spencerrivlin62,pope75,lundnovikov92},
\begin{equation}
\label{eq:tensorinvariants}
I_1 = \tr{ \LESS^2 },~~I_2 = \tr{ \LESOmega^2 },~~I_3 = \tr{ \LESS^3 },~~I_4 = \tr{ \LESS \LESOmega^2 },~~I_5 = \tr{ \LESS^2 \LESOmega^2 }.
\end{equation}
Examples of 
eddy viscosity and nonlinear models 
for large-eddy simulation
of the form of \cref{eq:modeddy,eq:modnonlin}
will be given in \cref{sec:modanalysis,sec:modexamples}.
In particular, 
in \cref{sec:modanalysis} we will analyze the behavior of existing subgrid-scale models with respect to the model constraints that 
will 
be discussed in \cref{sec:modconstraints}.
In \cref{sec:modexamples} we will show how these model constraints can lead to new subgrid-scale models.
In anticipation of the results we obtain there, we remark that 
the particular dependence of the 
functions $f_i$
on the tensor invariants of \cref{eq:tensorinvariants} plays a crucial role in determining a model's properties.

\section{Model constraints}
\label{sec:modconstraints}

As was alluded to in \cref{sec:intro}, the Navier-Stokes equations, \cref{eq:ns}, and the turbulent stresses, $\tau_{ij} = \filter{ u_i u_j } - \filter{ u }_i \filter{ u }_j$, have several interesting physical and mathematical properties.
One could argue that, to ensure physical consistency, it is desirable that these special properties are also exhibited by the equations of large-eddy simulation, \cref{eq:les}, and are not lost when modeling the turbulent stresses.
In what follows we will therefore provide a detailed discussion of several properties of the Navier-Stokes equations and the turbulent stresses.
We furthermore discuss the constraints that subgrid-scale models have to satisfy in order to preserve these properties.
In particular,
in \cref{sec:modconstraintssymm} we consider the 
symmetries
of the Navier-Stokes equations, whereas \cref{sec:modconstraintsnearwallscal} discusses the
desired
near-wall scaling behavior of the subgrid-scale stresses. 
Considerations relating to realizability 
are treated in \cref{sec:modconstraintsrealizability}.
Finally, several constraints on the 
production of subgrid-scale kinetic energy
are discussed in \cref{sec:modconstraintsSGSTKE}.

\subsection{Symmetry requirements}
\label{sec:modconstraintssymm}

The Navier-Stokes equations are 
% form 
invariant under several transformations of the coordinate system (see, \eg, Pope~\cite{pope11}).
As mentioned before,
these transformations, or symmetries, play an important 
physical
role because they ensure that the description of fluids is the same in all inertial frames of reference. 
They furthermore relate to conservation and scaling laws~\cite{razafindralandyetal07}.
Speziale~\cite{speziale85}, 
Oberlack~\cite{oberlack97,oberlack02} and
Razafindralandy \etal~\cite{razafindralandyetal07}
therefore
argue that
it is desirable that the basic equations of large-eddy simulation, 
\cref{eq:les},
admit the same symmetries as the 
Navier-Stokes equations, \cref{eq:ns}. 
This leads to a set of symmetry requirements for subgrid-scale models
that is discussed below. 
Let us first, however, provide more
detailed
information about the symmetries of the Navier-Stokes equations.

The 
unfiltered
incompressible
Navier-Stokes equations, \cref{eq:ns}, are 
% form
invariant under the following 
coordinate
transformations~\cite{pope11,oberlack97,oberlack02,razafindralandyetal07}:
%\begin{align}
%\label{eq:nssymmtimetrans}
%(t, x_i, u_i, p, \nu) & \rightarrow (t + T, x_i, u_i, p, \nu), \\
%\label{eq:nssymmprestrans}
%(t, x_i, u_i, p, \nu) & \rightarrow (t, x_i, u_i, p + P(t), \nu), \\
%\label{eq:nssymmgenGaltrans}
%(t, x_i, u_i, p, \nu) & \rightarrow (t, x_i + X_i(t), u_i + \dot{X}_i(t), p - \rho x_i \ddot{X}_i(t), \nu), \\
%\label{eq:nssymmorthtrans}
%(t, x_i, u_i, p, \nu) & \rightarrow (t, Q_{ij} x_j, Q_{ij} u_j, p, \nu), \\
%\label{eq:nssymmscaltrans}
%(t, x_i, u_i, p, \nu) & \rightarrow (e^{2a} t, e^{a + b} x_i, e^{-a + b} u_i, e^{-2a + 2b} p, e^{2b} \nu), \\
%\label{eq:nssymm2DMFI}
%(t, x_i, u_i, p, \nu) & \rightarrow (t, R_{ij}(t) x_j, R_{ij}(t)u_j + \dot{R}_{ij}(t)x_j, p + \tfrac{1}{2} \rho \omega_3^2 (x_1^2 + x_2^2) + 2 \rho \omega_3 \psi, \nu).
%\end{align}
\begin{itemize}
    \item \replace{}{the time translation,}
    \begin{equation}
    \label{eq:nssymmtimetrans}
    (t, x_i, u_i, p, \nu) \rightarrow (t + T, x_i, u_i, p, \nu);
    \end{equation}
    \item \replace{}{the pressure translation,}
    \begin{equation}
    \label{eq:nssymmprestrans}
    (t, x_i, u_i, p, \nu) \rightarrow (t, x_i, u_i, p + P(t), \nu);
    \end{equation}
    \item \replace{}{the generalized Galilean transformation,}
    \begin{equation}
    \label{eq:nssymmgenGaltrans}
    (t, x_i, u_i, p, \nu) \rightarrow (t, x_i + X_i(t), u_i + \dot{X}_i(t), p - \rho x_i \ddot{X}_i(t), \nu);
    \end{equation}
    \item \replace{}{orthogonal transformations,}
    \begin{equation}
    \label{eq:nssymmorthtrans}
    (t, x_i, u_i, p, \nu) \rightarrow (t, Q_{ij} x_j, Q_{ij} u_j, p, \nu);
    \end{equation}
    \item \replace{}{scaling transformations,}
    \begin{equation}
    \label{eq:nssymmscaltrans}
    (t, x_i, u_i, p, \nu) \rightarrow (e^{2a} t, e^{a + b} x_i, e^{-a + b} u_i, e^{-2a + 2b} p, e^{2b} \nu);
    \end{equation}
    \item \replace{}{and two-dimensional material frame-indifference,}
    \begin{equation}
    \label{eq:nssymm2DMFI}
    (t, x_i, u_i, p, \nu) \rightarrow (t, R_{ij}(t) x_j, R_{ij}(t)u_j + \dot{R}_{ij}(t)x_j, p + \tfrac{1}{2} \rho \omega_3^2 (x_1^2 + x_2^2) + 2 \rho \omega_3 \psi, \nu).
    \end{equation}
\end{itemize}
In the limit of an inviscid flow, $\nu \rightarrow 0$, the equations allow for an additional symmetry~\cite{oberlack02},
\begin{itemize}
    \item \replace{}{time reversal,}
    \begin{equation}
    \label{eq:eusymmtimerev}
    (t, x_i, u_i, p) \rightarrow (-t, x_i, -u_i, p).
    \end{equation}
\end{itemize}
In the time and pressure translations, \cref{eq:nssymmtimetrans,eq:nssymmprestrans}, $T$ and $P(t)$ indicate an arbitrary time shift and a time variation of the (background) pressure, respectively.
The generalized Galilean transformation, \cref{eq:nssymmgenGaltrans}, encompasses the space translation for $X_i(t)$ constant, and the classical Galilean transformation for $X_i(t)$ linear in time. 
Orthogonal transformations of the coordinate frame, \cref{eq:nssymmorthtrans}, are represented by a time-independent matrix $Q$ that is orthogonal, \ie, $Q_{ik} Q_{jk} = \delta_{ij}$.
These transformations correspond to instantaneous rotations and reflections of the coordinate system,
and include parity or spatial inversion~\cite{frisch95}.
The scaling transformations of \cref{eq:nssymmscaltrans} are parametrized by real $a$ and $b$. 
They originate from the fact that in mechanics arbitrary units can be used to measure space and time,
and they relate to the appearance of scaling laws, like \replace{to}{the} log law in wall-bounded flows~\cite{oberlack97,razafindralandyetal07}.
The transformation of \cref{eq:nssymm2DMFI} represents a time-dependent but constant-in-rate rotation of the coordinate system about the $x_3$ axis. It is characterized by a rotation matrix $R(t)$ with $\dot{R}_{ik} R_{jk} = \epsilon_{3ij}\omega_3$, 
for a constant rotation rate $\omega_3$. 
Here, $\epsilon_{ijk}$ represents the Levi-Civita symbol.
For the purposes of this transformation,
the flow is assumed to be confined to the $x_1$ and $x_2$ directions, so that $\psi$ represents the corresponding two-dimensional stream function. 
Invariance under this transformation is called material frame-indifference in the limit of a two-component flow,
\replace{sometimes}{also} referred to as two-dimensional material frame-indifference (2DMFI).
Refer to Oberlack~\cite{oberlack02} for more information about the interpretation of 2DMFI as an invariance (and not a material) property.
To avoid confusion, we remark that not all references provide the same expression for the transformed pressure\replace{.}{~\cite{oberlack97,oberlack02,razafindralandyetal07}.}
To the best of our knowledge, the expression we provide in \cref{eq:nssymm2DMFI}, which matches that of Oberlack~\cite{oberlack02}, is correct.
\replace{The final transformation, Eq.~(\ref{eq:eusymmtimerev}), will be referred to as time reversal.}{Finally, we consider the time reversal transformation, \cref{eq:eusymmtimerev}.}

To ensure physical consistency
with the 
Navier-Stokes
equations, \cref{eq:ns},
we will require
that the basic equations of large-eddy simulation,
\cref{eq:les},
are also invariant under the above symmetry transformations,
\cref{eq:nssymmtimetrans,eq:nssymmprestrans,eq:nssymmgenGaltrans,eq:nssymmorthtrans,eq:nssymmscaltrans,eq:nssymm2DMFI,eq:eusymmtimerev}.
Of course, we now have to read $\LESu_i$ instead of $u_i$ and $\LESp$ instead of $p$.
This results in the following symmetry requirements on the transformation behavior of the modeled subgrid-scale stresses~\cite{oberlack97}.
\begin{align}
\textrm{S1--3, S7: } \tauhatmod_{ij} &= \taumod_{ij}, \label{eq:symmreq1to3and7} \\
\textrm{S4: } \tauhatmod_{ij} &= Q_{im} Q_{jn} \taumod_{mn}, \label{eq:symmreq4} \\
\textrm{S5: } \tauhatmod_{ij} &= \mathrm{e}^{-2 a + 2 b} \taumod_{ij}, \label{eq:symmreq5} \\
\textrm{S6: } \tauhatmod_{ij} &= R_{im}(t) R_{jn}(t) \taumod_{mn}. \label{eq:symmreq6}
\end{align}
In symmetry requirements S1--3 and S7, \cref{eq:symmreq1to3and7}, the hat indicates application of 
the time or pressure translations, the generalized Galilean transformation, or time reversal, \cf~\cref{eq:nssymmtimetrans,eq:nssymmprestrans,eq:nssymmgenGaltrans,eq:eusymmtimerev}.
Symmetry requirements
S4 and S5
ensure invariance under instantaneous rotations and reflections, \cref{eq:nssymmorthtrans}, and scaling transformations, \cref{eq:nssymmscaltrans}, respectively.
Material frame-indifference in the limit of a two-component flow 
(invariance under \cref{eq:nssymm2DMFI})
holds when \cref{eq:symmreq6} is satisfied.

In the case of explicitly filtered large-eddy simulations, also the filtering operation needs to satisfy certain requirements to ensure that the above symmetry properties are not destroyed~\cite{oberlack97,razafindralandyetal07}.

\subsection{Near-wall scaling requirements}
\label{sec:modconstraintsnearwallscal}

Using 
numerical simulations, Chapman and Kuhn~\cite{chapmankuhn86} have revealed the near-wall scaling behavior of the time-averaged turbulent stresses. 
A simple 
model
for their observations can be obtained by performing a Taylor expansion of the velocity field in terms of the wall-normal distance~\cite{chapmankuhn86,sagaut06}.
As very close to a wall the tangential velocity components show a linear scaling with distance to that wall, the incompressibility constraint leads to a quadratic behavior for the wall-normal velocity.
From this the 
near-wall behavior of the time-averaged turbulent stresses can be derived.
Focusing on wall-resolved large-eddy simulations,
we would like to make sure that modeled stresses exhibit the same asymptotic behavior as the true turbulent stresses.
This ensures that, for instance, dissipative effects due to the model fall off quickly enough near 
solid boundaries.

In what follows, we will therefore require that modeled subgrid-scale stresses show the same near-wall behavior as the time-averaged true turbulent stresses, 
but then instantaneously. 
Denoting the wall-normal distance by $x_2$, we can express these
near-wall scaling requirements~(N) as
\begin{align}
\label{eq:modreqnearwallscal}
\begin{split}
\taumod_{11}, \taumod_{13}, \taumod_{33} &= \mathcal{O}(x_2^2), \\
\taumod_{12}, \taumod_{23} &= \mathcal{O}(x_2^3), \\
\taumod_{22} &= \mathcal{O}(x_2^4).
\end{split}
\end{align}

\subsection{Realizability requirements}
\label{sec:modconstraintsrealizability}

In the Reynolds-averaged Navier-Stokes (RANS) approach, 
instead of a spatial filtering operation, a time average is employed
to study the behavior of turbulent flows.
Consequently, 
in that approach,
the turbulent stress tensor 
is equal
to the Reynolds stress,
which represents a statistical average
and thus is symmetric positive semidefinite, 
also called realizable~\cite{duvachat77,schumann77}. 
\replace{Vreman, Geurts and Kuerten}{Vreman \etal}~\cite{vremanetal94} showed that, 
for positive spatial filters, 
the turbulent stress tensor of large-eddy simulation,
$\tau_{ij}$,
is also 
realizable.
They therefore argue that, from a theoretical point of view, it is desirable that subgrid-scale models exhibit realizability as well. 
A physical interpretation of realizability is given below.

Realizability of the turbulent stress tensor can be expressed in several equivalent ways~\cite{ghosal99}. 
For instance, it implies that the eigenvalues of $\tau_{ij}$, denoted by $k_1$, $k_2$ and $k_3$ here,
are nonnegative.
Consequently, the principal invariants of the turbulent stress tensor have to be nonnegative, as can be derived from their definition,
\begin{align}
\Ptau &= \tr{ \tau } = k_1 + k_2 + k_3, \\
\Qtau &= \frac{1}{2} ( \tr{ \tau }^2 - \tr{ \tau^2 } ) =  k_1 k_2 + k_2 k_3 + k_3 k_1, \\
\Rtau &= \det{ \tau } = k_1 k_2 k_3.
\end{align}
We will use $k = \frac{ 1 }{ 2 } \tr{ \tau }$ to denote the generalized subgrid-scale kinetic energy. 
The $k_i$ can therefore be interpreted as partial energies, which, from a physical point of view, preferably are positive.

When we separate the subgrid-scale model in an isotropic and a deviatoric part,
\begin{equation}
\taumod = \frac{ 2 }{ 3 } ( \kmod ) I + \taumoddev,
\end{equation}
realizability is guaranteed in case
\begin{align}
0 &\leq \Ptaumod = 2 \kmod, 
\label{eq:modreqrealizability1} \\
0 &\leq \Qtaumod  = \frac{ 4 }{ 3 } ( \kmod )^2 + \Qtaumoddev, \label{eq:modreqrealizability2} \\
0 &\leq \Rtaumod = \frac{ 8 }{ 27 } ( \kmod )^3 + \frac{ 2 }{ 3 } \kmod \Qtaumoddev + \Rtaumoddev, \label{eq:modreqrealizability3} \\
0 & \leq 4 ( -\Qtaumoddev )^3 - 27 ( \Rtaumoddev )^2. \label{eq:modreqrealizability4}
\end{align}
The last inequality ensures that the eigenvalues of $\taumod$ are real;
it is satisfied for all real symmetric $\taumoddev$.
Ordering the partial energies according to the definition $k_1 \geq k_2 = r k_1 \geq k_3 = s k_2 \geq 0$ and maximizing $\Rtau^2 / \Qtau^3$ and $\Qtau / \Ptau^2$ with respect to $s$ and $r$, we can further
obtain the following chain of inequalities~\cite{vreman04},
\begin{equation}
0 \leq \Rtaumod \leq \frac{ 1 }{ 3 \sqrt{3} } ( \Qtaumod )^{3/2} \leq \frac{ 1 }{ 27 } ( \Ptaumod )^3. \label{eq:modreqrealizability5}
\end{equation}
\cref{eq:modreqrealizability1,eq:modreqrealizability2,eq:modreqrealizability3,eq:modreqrealizability4,eq:modreqrealizability5} will be referred to as realizability conditions~(R) for the modeled subgrid-scale stresses.
The requirements of 
\cref{eq:modreqrealizability3,eq:modreqrealizability4} 
correspond to 
Lumley's triangle in the invariant map of the Reynolds stress anisotropy~\cite{lumleynewman77,lumley78}.
When no model is provided for the generalized subgrid-scale kinetic energy,
$\kmod$,
as is usually the case for eddy viscosity models,
useful bounds for this quantity can be obtained from the above inequalities~\cite{vremanetal94}.

\subsection{Requirements relating to the production of subgrid-scale kinetic energy}
\label{sec:modconstraintsSGSTKE}

In 
this
section 
we will look at energy transport in turbulent flows.
In particular, we 
will
focus on the transport of energy to small scales of motion
\replace{}{due to subgrid-scale models}, 
also referred to as 
subgrid
dissipation\replace{,}{}
or
the production of subgrid-scale kinetic energy.
Denoting the rate-of-strain tensor of the filtered velocity field by $\filter{ S } = S( \filter{ u })$, see \cref{eq:SandW},
we can express the 
true
subgrid dissipation as
\begin{equation}
\label{eq:subgriddissipation}
\Dtau = -\tr{ \tau \filter{S} }.\\
\end{equation}
The modeled subgrid dissipation, $\Dtaumod$, is defined analogously using $\taumod$ and $\LESS = \LESS( \LESu )$.
In \cref{sec:modconstraintsSGSTKEVreman} we will discuss Vreman's analysis of 
the actual subgrid dissipation, $\Dtau$,
along with his requirements for the modeled dissipation~\cite{vreman04}.
The requirements for the production of subgrid-scale kinetic energy of 
\replace{Nicoud, Baya Toda, Cabrit, Bose and Lee}{Nicoud \etal}~\cite{nicoudetal11}
are described in \cref{sec:modconstraintsSGSTKENicoud},
whereas the consequences of the second law of thermodynamics for subgrid-scale models are the topic of \cref{sec:modconstraintsSGSTKE2ndlaw}.
\cref{sec:modconstraintsSGSTKEVerstappen} treats Verstappen's minimum-dissipation condition for scale separation~\cite{verstappen11,verstappen16}.

\subsubsection{Vreman's model requirements}
\label{sec:modconstraintsSGSTKEVreman}

Vreman~\cite{vreman04} argues that
the turbulent stresses should be modeled
in such a way that the corresponding subgrid dissipation is small in laminar and transitional regions of a flow. 
On the other hand, the modeled 
dissipation should not be small where turbulence occurs.
This
ensures
that subgrid-scale models are neither overly, nor underly dissipative, thereby preventing unphysical transition
from a laminar to a turbulent flow 
and vice versa
in a large-eddy simulation.

To realize the above situation,
Vreman requires that the modeled production of subgrid-scale kinetic energy vanishes for flows for which the actual production is known to be zero.
If, on the other hand, for a certain flow it is known that there \emph{is} energy transport to
subgrid scales, the model should show the same behavior. 
Vreman's model requirements
for the production of subgrid-scale kinetic energy
can be summarized in the following form:
\begin{alignat}{2}
\label{eq:modreqVremana}
\textrm{P1a: } \Dtaumod & = 0    & \textrm{ when } \Dtau & = 0, \\
\label{eq:modreqVremanb}
\textrm{P1b: } \Dtaumod & \neq 0 & \textrm{ when } \Dtau & \neq 0.
\end{alignat}
To study the behavior of both the actual and the modeled subgrid dissipation,
Vreman developed a classification of flows based on the number and position of zero elements in the (unfiltered) velocity gradient tensor. 
A total of 320 flow types can be distinguished, corresponding to all incompressible velocity gradients having zero to nine vanishing elements. 
Nonzero elements are left unspecified. 
Vreman shows that, for general filters, there are only thirteen flow types for which the true subgrid dissipation, $\Dtau$, always vanishes. 
He calls such flow types locally laminar and refers to their collection as the flow algebra of $\Dtau$. 
Assuming the use of an isotropic filter to compute the true subgrid dissipation, 
we include three more flow classes in this set.
It can be shown that the true subgrid dissipation is not generally zero for any of the other 304 flow classes.
Note that there exist two-component flows that belong to these latter classes
and that, therefore, they do not necessarily have a zero subgrid dissipation\replace{}{ (in contrast to what is required in \cref{sec:modconstraintsSGSTKENicoud})}.

Using Vreman's classification of flows, we 
thus
obtain sixteen flow types for which we would like the modeled subgrid dissipation to vanish and 304 flow types for which, preferably, $\Dtaumod$ is not generally zero.
Although specific flows may exist that show a different behavior, we will in fact consider P1a to be fulfilled when $\Dtaumod$ vanishes for the sixteen laminar flow types, 
and P1b when the modeled subgrid dissipation is nonzero for the remaining (nonlaminar) flow types.

\subsubsection[Nicoud et al. model requirements]{Nicoud \etal~model requirements}
\label{sec:modconstraintsSGSTKENicoud}

On the basis of physical grounds, \replace{Nicoud, Baya Toda, Cabrit, Bose and Lee }{Nicoud \etal}~\cite{nicoudetal11} argue that certain flows cannot be maintained if energy is transported to subgrid scales. 
They therefore see it as a desirable property that the modeled production of subgrid-scale kinetic energy vanishes for these flows.
In particular, they require that a subgrid-scale model is constructed in such a way that the subgrid dissipation is zero for all two-component flows~(P2a) and for the pure axisymmetric strain~(P2b).

It should be noted that 
requirement P2a
cannot be reconciled with Vreman's 
second 
model constraint 
(\hyperref[eq:modreqVremanb]{P1b}),
as the latter requires that certain two-component 
flows have a nonzero subgrid dissipation.
Apparently, the physical reasoning employed by Nicoud \etal~\cite{nicoudetal11} is not compatible with the mathematical properties of the turbulent stress tensor that were discovered by Vreman~\cite{vreman04}. 
For comparison we will, however, not exclude any requirements in what follows.

\subsubsection{Consistency with the second law of thermodynamics}
\label{sec:modconstraintsSGSTKE2ndlaw}

In turbulent flows, energy can be transported from large to small scales (forward scatter) and vice versa (backscatter). 
\replace{}{As mentioned in \cref{sec:LESmodels}, the net transport of energy, which is from large to small scales of motion, is often parametrized using dissipative subgrid-scale models.}
%such as eddy viscosity models
The second law of thermodynamics requires that the \replace{net transport of energy is of the former type}{total dissipation in flows is nonnegative}~\cite{razafindralandyetal07}.
Assuming that, apart from the subgrid dissipation, only viscous dissipation plays a role in large-eddy simulation, we thus need,
\begin{equation}
\label{eq:modreq2ndlaw}
\textrm{P3: } \Dtaumod + 2 \nu I_1 \geq 0.
\end{equation}
\replace{}{The viscous dissipation, $2 \nu I_1 = 2 \nu \tr{ \LESS^2 }$, is a positive quantity.
The second law therefore allows the production of subgrid-scale kinetic energy, $\Dtaumod$, to become negative.}
In a practical large-eddy simulation, subgrid-scale motions \replace{may not be}{are often not} well resolved.
\replace{In that case, one}{One} could \replace{}{therefore} argue that \replace{backscatter should be excluded completely}{a negative production of subgrid-scale kinetic energy due to subgrid-scale models should be precluded}, to prevent numerical errors from growing to the size of large-scale motions.
\replace{Backscatter due to subgrid-scale models is prevented when the modeled subgrid dissipation, $\Dtaumod$, is nonnegative.}{}
\replace{That is}{To that end}, one can simply drop
the second term on the left-hand side of \cref{eq:modreq2ndlaw}\replace{,
because
from the definition of the tensor invariants, Eq.~(\ref{eq:tensorinvariants}), it follows that $2 \nu I_1 = 2 \nu \tr{ \LESS^2 } \geq 0$.}{.}
\replace{}{%Do note that, if one only employs an eddy viscosity term with a nonnegative subgrid dissipation, backscatter cannot be represented. 
Do note that subgrid-scale models consisting of only an eddy viscosity term with a nonnegative subgrid dissipation cannot capture backscatter.
Additional model terms, such as (nondissipative) tensor terms that are nonlinear in the velocity gradient, would be required for that purpose.}

\subsubsection{Verstappen's model requirements}
\label{sec:modconstraintsSGSTKEVerstappen}

When the filtered Navier-Stokes equations, \cref{eq:filtns}, are supplied with a subgrid-scale model, 
one obtains a closed set of equations for the large-scale velocity field, given by \cref{eq:les}.
Solutions of \cref{eq:les}, however, are not necessarily independent of scales of motion smaller than the 
filter width, $\filterlength$.
Indeed, due to the convective nonlinearity, energy transport takes place between large and small scales of motion.
This is 
troublesome when the small scales of motion
are not well resolved,
as is commonly the case in numerical simulations.
Verstappen~\cite{verstappen11,verstappen16} therefore argues that subgrid-scale models should be constructed in such a way that the basic equations of large-eddy simulation, \cref{eq:les}, 
provide a
solution of large-scale dynamics, independent of small-scale motions. 
Stated otherwise,
subgrid-scale models have to cause scale separation.
This can be achieved by ensuring that subgrid-scale models counterbalance the convective production of small-scale kinetic energy
and dissipate any kinetic energy (initially) contained in small scales of motion.

It can be shown that the kinetic energy of subgrid-scale motions is influenced by both large and small scales of motions.
Because the behavior of the small scales of motion is not fully known in a large-eddy simulation,
this complicates the construction of subgrid-scale models that dissipate
this energy.
We can, however, apply
Poincaré's inequality to bound the kinetic energy of
small-scale motions
in terms of
the magnitude of the 
velocity gradient.
To that end, anticipating discretization using a finite-volume method, we divide the flow domain into a number of small non-overlapping (control) volumes,
characterized by a length scale $\testfilterlength \ge \filterlength$.
Here, $\delta$ is the subgrid characteristic length scale (or filter length), commonly associated with the mesh size.
Further defining $\testfilter{ \cdot }$ as the average over a volume $V_{ \testfilterlength }$, we can write Poincaré's inequality for the small-scale kinetic energy contained in this volume as
\begin{equation}
\label{eq:PoincareSFSkineticenergy}
\int_{ V_{ \testfilterlength } } \frac{ 1 }{ 2 } ( \LESu_i - \testfilter{ \LESu }_i )( \LESu_i - \testfilter{ \LESu }_i ) \dif V 
\le 
C_{ \testfilterlength } \int_{ V_{ \testfilterlength } } \frac{ 1 }{ 2 } \frac{ \partial \LESu_i }{ \partial x_j } \frac{ \partial \LESu_i }{ \partial x_j } \dif V.
\end{equation}
Here, $C_{ \testfilterlength }$ is called Poincaré's constant, which depends only on the filter volume, $V_{ \testfilterlength }$.
We can now render motions that are smaller than the 
length scale 
$\testfilterlength$ inactive by forcing the right-hand side of \cref{eq:PoincareSFSkineticenergy} to zero.
That is,
we need
\begin{align}
\label{eq:modreqVerstappen}
\begin{split}
\textrm{P4: } 0 > \frac{ \dif }{ \dif t } & \int_{ V_{ \testfilterlength } } \frac{ 1 }{ 2 } \frac{ \partial \LESu_i }{ \partial x_j } \frac{ \partial \LESu_i }{ \partial x_j } \dif V = 
\frac{ \dif }{ \dif t } \int_{ V_{ \testfilterlength } } \frac{ 1 }{ 2 } (I_1 - I_2) \dif V = \\
& \int_{ S_{ \testfilterlength } } 
[ 
- \frac{ 1 }{ 2 } (I_1 - I_2) \LESu_k 
- \frac{ 1 }{ \rho } \frac{ \partial ( \LESp + \frac{ 2 }{ 3 } \rho \kmod ) }{ \partial x_j } \frac{ \partial \LESu_k }{ \partial x_j } 
+ \nu \frac{ \partial^2 \LESu_i }{ \partial x_j \partial x_j } \frac{ \partial \LESu_i }{ \partial x_k }
- \frac{ \partial \taumoddev_{ij} }{ \partial x_j } \frac{ \partial \LESu_i }{ \partial x_k }
]
n_k \dif S \\
+ & \int_{ V_{ \testfilterlength } } 
[
- ( I_3 - I_4 )
- \nu \frac{ \partial^2 \LESu_i }{ \partial x_j \partial x_j } \frac{ \partial^2 \LESu_i }{ \partial x_k \partial x_k }
+ \frac{ \partial \taumoddev_{ij} }{ \partial x_j } \frac{ \partial^2 \LESu_i }{ \partial x_k \partial x_k }
] \dif V.
\end{split}
\end{align}
Here, the evolution of (half) the squared velocity gradient magnitude,
$\frac{ 1 }{ 2 } ( I_1 - I_2) = \frac{ 1 }{ 2 } ( \tr{ \LESS^2 } - \tr{ \LESOmega^2 } )$,
in a 
volume
of size $V_{ \testfilterlength }$ is expressed as a sum of two contributions, the one being a surface integral that relates to transport processes, and the other a volume integral, that represents body forces.
The quantity $-( I_3 - I_4 ) = -( \tr{ \LESS^3 } - \tr{ \LESS \LESOmega^2 } )$
represents the convective production of velocity gradient.
If transport processes are ignored, it is this production that
has to be counterbalanced by the subgrid-scale model to 
ensure that small-scale motions disappear.

\section{Analysis of existing subgrid-scale models}
\label{sec:modanalysis}

Before illustrating how new subgrid-scale models can be constructed using the constraints that were discussed in \cref{sec:modconstraints},
let us analyze the properties of existing models.
We will focus on the properties of subgrid-scale models that 
depend locally on the
velocity gradient, as introduced in \cref{sec:LESmodels}.

\subsection{Some existing subgrid-scale models}
\label{sec:existingmods}

In large-eddy simulation often eddy viscosity models, \cref{eq:modeddy},
are employed to parametrize the effects of small-scale motions on turbulent flows.
There are several eddy viscosity models that
depend 
on the velocity gradient 
in a local and isotropic fashion.
That is, 
they do not involve additional transport equations, and
they are invariant under rotations of the coordinate system 
(refer to property \hyperref[eq:symmreq4]{S4} of \cref{eq:symmreq4}).
These models can therefore be expressed using 
the tensor invariants of \cref{eq:tensorinvariants}, provided again for convenience,
\begin{equation}
\label{eq:tensorinvariants2}
I_1 = \tr{ \LESS^2 },~~I_2 = \tr{ \LESOmega^2 },~~I_3 = \tr{ \LESS^3 },~~I_4 = \tr{ \LESS \LESOmega^2 },~~I_5 = \tr{ \LESS^2 \LESOmega^2 }.
\end{equation}
We have, for example,
\begin{alignat}{2}
\label{eq:modSmag}
%TODO Add \displaybreak if big block of equations needs to flow over to the next page
%\displaybreak \\
& \textrm{Smagorinsky~\cite{smagorinsky63}:}  &  \nuesup{S} & = ( \Csub{S} \filterlength )^2 \sqrt{ 2 I_1 }, \\ 
\label{eq:modWALE}
& \textrm{WALE~\cite{nicoudducros99}:}  &  \nuesup{W} & = ( \Csub{W} \filterlength )^2 \frac{ J^{3/2} }{ I_1^{5/2} + J^{5/4} },
\textrm{ where } J = \frac{ 1 }{ 6 } ( I_1 + I_2 )^2 + 2(I_5 - \frac{ 1 }{ 2 } I_1 I_2), \\
\label{eq:modVreman}
& \textrm{Vreman~\cite{vreman04}:}  &  \nuesup{V} & = ( \Csub{V} \filterlength )^2 \sqrt{ \frac{ \QGGT }{ \PGGT } }, \\
\label{eq:modsigma}
& \sigma\textrm{~\cite{nicoudetal11}:}  &  \nuesup{\sigma} & = ( \Csub{\sigma} \filterlength )^2 \frac{ \sigma_3 ( \sigma_1 - \sigma_2 ) ( \sigma_2 - \sigma_3 ) }{ \sigma_1^2 }, \\
\label{eq:modQR}
& \textrm{QR~\cite{verstappenetal10,verstappen11,verstappenetal14}:}  &  \nuesup{QR} & = ( \Csub{QR} \filterlength )^2 \frac{ \max\{0, - I_3\} }{ I_1 }, \\
\label{eq:modS3PQR}
& \textrm{S3PQR~\cite{triasetal15}:}  &  \nuesup{S3} & = ( \Csub{S3} \filterlength )^2 \PGGT^p \QGGT^{-( p + 1 )} \RGGT^{ ( p + 5/2 )/3 },\\
\label{eq:modAMD}
& \textrm{AMD~\cite{rozemaetal15}:}  &  \nuesup{A} & = ( \Csub{A} \filterlength )^2 \frac{ \max\{ 0, -( I_3 - I_4 )\} }{ I_1 - I_2 }.
\end{alignat}
Here, the $C$'s are used to denote model constants, whereas 
$\filterlength$ represents the subgrid characteristic length scale (or filter length) of the large-eddy simulation.
In \cref{eq:modVreman,eq:modS3PQR}, the quantities
\begin{align}
\label{eq:GGTPQR}
\begin{alignedat}{3}
\PGGT & = I_1 - I_2, \qquad & 
\QGGT & = \frac{ 1 }{ 4 } ( I_1 + I_2 )^2 + 4 (I_5 - \frac{ 1 }{ 2 } I_1 I_2 ), \qquad & 
\RGGT & = \frac{ 1 }{ 9 } ( I_3 + 3 I_4 )^2,
\end{alignedat}
\end{align}
are the tensor invariants of 
$GG^T = \LESS^2 - \LESOmega^2 - (\LESS \LESOmega - \LESOmega \LESS)$.
The $\sigma_i$ in \cref{eq:modsigma} represent the square roots of the eigenvalues of this 
same
tensor or, equivalently,
the singular values of the velocity gradient, $G$, \cref{eq:velgrad},
with $\sigma_1 \geq \sigma_2 \geq \sigma_3 \geq 0$.
Their expression in terms of the invariants of \cref{eq:GGTPQR} can be found in Nicoud \etal~\cite{nicoudetal11}.
To avoid confusion, it is to be noted that the Q and R in the name of the QR model refer to the second and third invariants of the rate-of-strain tensor, 
$\QS = -\frac{ 1 }{ 2 } I_1 = -\frac{ 1 }{ 2 } \tr{ \LESS^2 }$  and $\RS = \frac{ 1 }{ 3 } I_3 = \frac{ 1 }{ 3 } \tr{ \LESS^3 }$, respectively.
The S3PQR model essentially forms a class of models, one for each value of the parameter $p$.
Following the work by Trias \etal~\cite{triasetal15},
we will in particular discuss 
the S3PQ ($p = -5/2$), S3PR ($p = -1$) and S3QR ($p = 0$) models.
Finally, note that \cref{eq:modVreman,eq:modAMD} provide isotropized versions of Vreman's model~\cite{vreman04} and the anisotropic minimum-dissipation (AMD) model of Rozema
\etal~\cite{rozemaetal15}, respectively.

A specific example of a nonlinear model of the form of \cref{eq:modnonlin} is given by the gradient
model~\cite{leonard74,clarketal79},
\begin{align}
\label{eq:modgrad}
\begin{split}
\taumodsub{G} & = \Csub{G} \filterlength^2 \left( \LESS^2 - \LESOmega^2 - (\LESS\LESOmega - \LESOmega\LESS)\right).
\end{split}
\end{align}
A different nonlinear model is the explicit algebraic subgrid-scale stress model (EASSM) of Marstorp \etal~\cite{marstorpetal09}, which 
in its 
nondynamic
version can be written
\begin{equation}
\label{eq:modEASSM}
\taumodsub{E} = \Csub{E} \filterlength^2 \frac{ f }{ f^2 - I_2 / I_1 } \left( \frac{ 80 }{ 99 }  I_1 I - \sqrt{ 2 I_1 } \frac{ f }{ f^2 - I_2 / I_1  } \LESS - \frac{ 1 }{ f^2 - I_2 / I_1 } ( \LESS \LESOmega - \LESOmega \LESS ) \right).
\end{equation}
Here, 
$f = f(I_2 / I_1)$ is a function that tends to 1 as $I_2 / I_1$ goes to 0. 
It corresponds to $9 c_1 / ( 4 c_3 )$ in the notation of Marstorp \etal~\cite{marstorpetal09}.

\cref{tab:modanalysis} provides a summary of the behavior of the above subgrid-scale models with respect to the model requirements discussed in \cref{sec:modconstraints}.
A detailed
discussion of these results is presented in what follows.

\begin{table}[!t]
    \caption{
        \label{tab:modanalysis}
        Summary of the properties of several subgrid-scale models. 
        The properties considered are 
        \hyperref[eq:symmreq1to3and7]{S1--4}:~time, pressure, generalized Galilean, and rotation and reflection invariance;
        \hyperref[eq:symmreq5]{S5}:~scaling invariance;
        \hyperref[eq:symmreq6]{S6}:~two-dimensional material frame-indifference;
        \hyperref[eq:symmreq1to3and7]{S7}:~time reversal invariance;
        \hyperref[eq:modreqnearwallscal]{N}:~the proper near-wall scaling behavior;
        \hyperref[eq:modreqrealizability1]{R}:~realizability;
        \hyperref[eq:modreqVremana]{P1a}:~zero subgrid dissipation for laminar flow types;
        \hyperref[eq:modreqVremanb]{P1b}:~nonzero subgrid dissipation for nonlaminar flow types;
        \hyperref[sec:modconstraintsSGSTKENicoud]{P2a}:~zero subgrid dissipation for two-component flows;
        \hyperref[sec:modconstraintsSGSTKENicoud]{P2\replace{a}{b}}:~zero subgrid dissipation for the pure axisymmetric strain;
        \hyperref[eq:modreq2ndlaw]{P3}:~consistency with the second law of thermodynamics;
        \hyperref[eq:modreqVerstappen]{P4}:~sufficient subgrid dissipation for scale separation.
        The horizontal rule separates eddy viscosity models from models that are nonlinear in the velocity gradient\replace{}{.}
    }
    \begin{tabular}{lccccccccccccc}
        \toprule
        ~                ~                                                     & Eq.                      & S1--4 & S5* &  S6  &  S7* & N* & R &  P1a &  P1b &  P2a & P2b &  P3  &  P4  \\
        \midrule
        Smagorinsky      \cite{smagorinsky63}                                  & \labelcref{eq:modSmag}   &   Y   &  N  &   Y  &   N  &  N & ~ &   N  &  Y   &   N  &  N  &   Y  &   Y  \\
        WALE             \cite{nicoudducros99}                                 & \labelcref{eq:modWALE}   &   Y   &  N  &   N  &   N  &  Y & ~ &   N  &  Y   &   N  &  N  &   Y  &   Y  \\
        Vreman           \cite{vreman04}                                       & \labelcref{eq:modVreman} &   Y   &  N  &   N  &   N  &  N & ~ &   N  &  Y   &   N  &  N  &   Y  &   Y  \\
        $\sigma$         \cite{nicoudetal11}                                   & \labelcref{eq:modsigma}  &   Y   &  N  &   Y  &   N  &  Y & ~ &   Y  &  N   &   Y  &  Y  &   Y  &   N  \\
        QR               \cite{verstappenetal10,verstappen11,verstappenetal14} & \labelcref{eq:modQR}     &   Y   &  N  &   Y  &   N  &  N & ~ &   Y  &  N   &   Y  &  N  &   Y  &   N  \\
        S3PQR            \cite{triasetal15}                                    & \labelcref{eq:modS3PQR}  &   Y   &  N  & Y\dg & N\dg &  Y & ~ & Y\dg & Y\dg & Y\dg &  N  & Y\dg & Y\dg \\
        - S3PQ           ~                                                     & ~                        &   Y   &  N  &   N  &   N  &  Y & ~ &   N  &  Y   &   N  &  N  &   Y  &   Y  \\
        - S3PR           ~                                                     & ~                        &   Y   &  N  &   Y  & N\dg &  Y & ~ &   Y  &  N   &   Y  &  N  & Y\dg &   N  \\
        - S3QR           ~                                                     & ~                        &   Y   &  N  &   Y  & N\dg &  Y & ~ &   Y  &  N   &   Y  &  N  & Y\dg &   N  \\
        AMD              \cite{rozemaetal15}                                   & \labelcref{eq:modAMD}    &   Y   &  N  &   Y  &   N  &  N & ~ &   Y  &  N   &   Y  &  N  &   Y  &   Y  \\
        Vortex stretching ~                                                    & \labelcref{eq:modVS}     &   Y   &  N  &   Y  &   N  &  Y & ~ &   Y  &  N   &   Y  &  Y  &   Y  &   N  \\
        \midrule
        Gradient         \cite{leonard74,clarketal79}                          & \labelcref{eq:modgrad}   &   Y   &  N  &   N  &   Y  &  N & Y &   Y  &  N   &   Y  &  N  &   N  &   ~  \\
        EASSM            \cite{marstorpetal09}                                 & \labelcref{eq:modEASSM}  &   Y   &  N  &   N  &   N  &  N & Y &   N  &  Y   &   N  &  N  &   Y  &   ~  \\ % Nondynamic version
        Nonlinear example ~                                                    & \labelcref{eq:modNL}     &   Y   &  N  &   Y  &   Y  &  Y & N &   Y  &  N   &   Y  &  N  &   N  &   ~  \\
        \bottomrule
    \end{tabular}
    *~The dynamic procedure~\cite{germanoetal91} may restore these properties~\cite{caratietal01,oberlack97,razafindralandyetal07}.
    \\
    \dg~Depending on the value of the model parameter\replace{}{, $p$,} and/or the implementation.
\end{table}

\subsection{Discussion of symmetry properties}
\label{sec:modanalysissymm}

It can be shown that
time~(\hyperref[eq:symmreq1to3and7]{S1}), pressure~(\hyperref[eq:symmreq1to3and7]{S2}) and generalized Galilean~(\hyperref[eq:symmreq1to3and7]{S3}) invariance are automatically satisfied by subgrid-scale models that are based on the velocity gradient alone, \cref{eq:modnonlin}.
Furthermore, such models satisfy rotation and reflection invariance~(\hyperref[eq:symmreq4]{S4}) when their model coefficients, $\alpha_i$, only depend on the velocity gradient via the tensor invariants of 
\cref{eq:tensorinvariants2}.
All the existing models discussed above are of this form
and thus satisfy these symmetries.

Invariance under scaling transformations~(\hyperref[eq:symmreq5]{S5})
is not straightforwardly satisfied, because it requires an intrinsic length scale, that is, a length scale that is directly related to the properties of a flow~\cite{oberlack97,razafindralandyetal07}.
Neither the velocity gradient, which has units of inverse time, nor the externally imposed 
length scale, $\filterlength$, 
which is usually related to the mesh size in a numerical simulation,
can provide this.
Therefore, none of the models listed before satisfies scale invariance out of itself.
As a consequence, simulations using these models can in principle not capture certain scaling laws, like the well-known log law of wall-bounded flows, or certain self-similar solutions~\cite{oberlack97,razafindralandyetal07}.
If model constants
are determined dynamically~\cite{germanoetal91}
in a numerical simulation,
scale invariance is known to be restored~\cite{oberlack97,razafindralandyetal07}. 
Application
of the dynamic procedure is beyond the scope of the current study, 
but let us note that it relies on an explicit filtering operation.
This may destroy some 
of the symmetries of the Navier-Stokes equations unless certain
restrictions on the filter are fulfilled~\cite{oberlack97,razafindralandyetal07}.

With respect to material frame-indifference in the limit of a two-component flow~(2DMFI, \hyperref[eq:symmreq6]{S6}) we remark that two-component flows can be characterized by the following set of invariants 
(\cf~\cref{eq:tensorinvariants2}):
\begin{align}
\label{eq:2Cflowchar}
I_3 = I_4 = I_5 - \frac{ 1 }{ 2 } I_1 I_2 = 0.
\end{align}
Therefore, eddy viscosity models that are based on these quantities are 2DMFI.
Also $I_1$ and $I_5 / I_2$ are 2DMFI quantities and may appear.
$I_2$ cannot appear by itself without violating 2DMFI.
We believe, however, that nonlinear model terms that involve the rate-of-rotation tensor do not necessarily have to be discarded to make a subgrid-scale model compatible with 2DMFI,
as long as the coefficients of such terms vanish in two-component flows.

Although the turbulent stress tensor is invariant under time reversal~(\hyperref[eq:symmreq1to3and7]{S7}),
time reversibility is not generally regarded as a desirable property of subgrid-scale models.
Indeed, \replace{Carati, Winckelmans and Jeanmart}{Carati \etal}~\cite{caratietal01} argue that at least one of the terms comprising a subgrid-scale model has to lead to an irreversible loss of information.
Most of the listed eddy viscosities are positive for all possible flow fields. 
This ensures irreversibility as well as consistency with the second law of thermodynamics~(\hyperref[eq:modreq2ndlaw]{P3}).
In contrast, the gradient model is time reversal invariant as a whole,
while
the explicit algebraic subgrid-scale stress model has the interesting property that the term linear in $\LESS$ is not time reversal invariant,
whereas both other terms are.
It is to be noted that the dynamic procedure~\cite{germanoetal91} can restore time reversal invariance of the modeled subgrid-scale stresses when it is used without clipping.
In the light of the above discussion, this, however, is sometimes seen as an artifact of this method~\cite{caratietal01}.

\subsection{Discussion of the near-wall scaling behavior}
\label{sec:modanalysisnearwallscal}

The near-wall scaling behavior of the different subgrid-scale models can readily be deduced using the asymptotic analysis described in \cref{sec:modconstraintsnearwallscal}.
For instance,
we find that
$I_1, I_2, I_5 = \mathcal{O}( 1 )$ and $I_3, I_4 = \mathcal{O}( x_2 )$, whereas
certain special combinations of these quantities show a different scaling,
namely
$I_1 + I_2 = \mathcal{O}( x_2^2 )$, $I_3 + 3 I_4 = \mathcal{O}( x_2^3 )$ and $I_5 - \frac{ 1 }{ 2 } I_1 I_2 = \mathcal{O}( x_2^2 )$. 
From this we can determine the near-wall asymptotic behavior of each of the model coefficients, that can subsequently be compared to the desired behavior.
The desired scaling behavior of the eddy viscosity is $\nue = \mathcal{O}( x_2^3 )$.
Some subgrid-scale models automatically exhibit this behavior and, therefore, make sure that dissipative effects are not too prominent near a wall.
Damping functions or the dynamic procedure can be used to correct the near-wall behavior of 
the other
subgrid-scale models~\cite{sagaut06}.

\subsection{Discussion of realizability}
\label{sec:modanalysisrealizability}

As remarked in \cref{sec:modconstraintsrealizability},
to decide on the realizability of a subgrid-scale model, it needs to include a model for the generalized subgrid-scale kinetic energy.
For the aforementioned eddy viscosity models such a model is not supplied,
leaving it to a modified pressure term.
Therefore we cannot assess these eddy viscosity models based on realizability requirements~(\hyperref[eq:modreqrealizability1]{R}).
One could, however, use \cref{eq:modreqrealizability1,eq:modreqrealizability2,eq:modreqrealizability3,eq:modreqrealizability4,eq:modreqrealizability5} to find a model for the generalized subgrid-scale kinetic energy that,
when added to an eddy viscosity model,
provides a realizable subgrid-scale model.

The gradient model and the explicit algebraic subgrid-scale stress model both provide an explicit model for the subgrid-scale kinetic energy and can be shown to be realizable.

\subsection{Discussion of the production of subgrid-scale kinetic energy}
\label{sec:modanalysisSGSTKE}

The production of subgrid-scale kinetic energy due to eddy viscosity models, \cref{eq:modeddy}, can be expressed as
\begin{equation}
\label{eq:modeddydiss}
\Dmodsub{e} = 2 \nue I_1.
\end{equation}
The behavior of this quantity is mostly determined by $\nue$, as $I_1 = \tr{ \LESS^2 }$ is nonnegative and only vanishes in purely rotational flows.
For the gradient model, \cref{eq:modgrad}, we have
\begin{equation}
\label{eq:modgraddiss}
\Dmodsub{G} = \Csub{G} \filterlength^2 ( I_3 - I_4 ).
\end{equation}
This quantity does not have a definite sign, since $I_3 - I_4 = \tr{ \LESS^3 } - \tr{ \LESS \LESOmega^2 }$.
The subgrid dissipation of the explicit algebraic subgrid-scale stress model, \cref{eq:modEASSM}, is 
entirely due to the term linear in the rate-of-strain tensor, as the other terms are orthogonal to it. Therefore, $\Dmodsub{E}$ has an expression similar to \cref{eq:modeddydiss}.

As discussed in \cref{sec:modconstraintsSGSTKEVreman}, the dissipation behavior of subgrid-scale models can be studied using Vreman's classification of flows~\cite{vreman04}.
In particular,
we can determine the flow algebra of a model's subgrid dissipation, \ie, the set of flows for which this dissipation vanishes.
Subsequently, we can compare it to the flow algebra of the subgrid dissipation of the true turbulent stresses.
This is done in \cref{tab:modconstraintsVreman}, which provides
a summary of the size of the flow algebra of different quantities, including the subgrid-scale models discussed above.
To determine whether a subgrid-scale model satisfies Vreman's model requirements~(\hyperref[eq:modreqVremana]{P1a}, \hyperref[eq:modreqVremanb]{P1b}),
its flow algebra can be compared to the desired outcome, listed next to $\Dtau$.
Not a single 
model
was found with exactly the same dissipation behavior as the true subgrid dissipation.
This contrasts with Vreman's findings, which is due to the fact that,
in computing the true turbulent stresses, $\tau_{ij}$,
we assumed the use of a filter that conforms to the symmetry properties of the Navier-Stokes equations and, thus, is isotropic.
Eddy-viscosity-type models that are constructed using quantities that have a smaller flow algebra than the actual subgrid dissipation can be expected to be too dissipative. On the other hand, a model based on a quantity that is zero more often than $\Dtau$, can be expected to be underly dissipative.

\begin{table}[!t]
    \centering
    \caption{
        \label{tab:modconstraintsVreman}
        Summary of the size of the flow algebra of the true subgrid dissipation $\Dtau$, \cref{eq:subgriddissipation}, and of several quantities based on the tensor invariants of
        \cref{eq:tensorinvariants2}. 
        $Q_n$ represents the set of flow types for which the velocity gradient contains $n$ zero elements. 
        The total number of flows in Vreman's classification (3D), and the number of two-component (2C) flows, \cref{eq:2Cflowchar},
        are listed for reference. 
        Results provided here differ slightly from those of Vreman~\cite{vreman04} because we assumed the use of an isotropic filter to compute $\Dtau$.
    }
    \begin{tabular} {lccccccccccc}
        \toprule
        ~                                                            & $Q_0$ & $Q_1$ & $Q_2$ & $Q_3$ & $Q_4$ & $Q_5$ & $Q_6$ & $Q_7$ & $Q_8$ & $Q_9$ & $Q_{0-9}$ \\
        \midrule
        3D flows                                                     &    1  &    9  &   33  &   66  &   81  &   66  &   39  &   18  &    6  &    1  &    320    \\
        2C flows                                                     &       &       &       &       &       &    3  &    6  &   12  &    6  &    1  &     28    \\
        $\Dtau$                                                      &       &       &       &       &       &       &       &    9  &    6  &    1  &     16    \\
        \midrule
        $I_1$, $\nuesup{S}$, $\Dmodsub{E}$                           &       &       &       &       &       &       &       &       &       &    1  &      1    \\
        $\PGGT$                                                      &       &       &       &       &       &       &       &       &       &    1  &      1    \\
        %       $I_2$                                                        &       &       &       &       &       &       &    1  &    3  &       &    1  &      5    \\
        %       $I_5$                                                        &       &       &       &       &       &       &    1  &    3  &       &    1  &      5    \\
        $\QGGT$, $\nuesup{W}$, $\nuesup{V}$, $\nuesup{S3PQ}$         &       &       &       &       &       &       &       &    6  &    6  &    1  &     13    \\
        $I_1 + I_2$                                                  &       &       &       &       &       &       &    8  &   12  &    6  &    1  &     27    \\
        $I_5 - \frac{ 1 }{ 2 } I_1 I_2$, $\nuesup{VS}$               &       &       &       &       &       &    3  &    7  &   12  &    6  &    1  &     29    \\
        $I_3$, $\nuesup{QR}$, $\Dmodsub{NL}$                         &       &       &       &       &       &    6  &   18  &   18  &    6  &    1  &     49    \\
        %       $I_4$                                                        &       &       &       &       &       &    6  &   19  &   18  &    6  &    1  &     50    \\
        $\nuesup{A}$, $\Dmodsub{G}$                                  &       &       &       &       &       &    6  &   20  &   18  &    6  &    1  &     51    \\
        %       $I_6$                                                        &       &       &       &       &    3  &   15  &   19  &   12  &    6  &    1  &     56    \\
        $\RGGT$, $\nuesup{\sigma}$, $\nuesup{S3PR}$, $\nuesup{S3QR}$ &       &       &       &    6  &   30  &   48  &   36  &   18  &    6  &    1  &    145    \\
        \bottomrule
    \end{tabular}
\end{table}

\Cref{tab:modconstraintsVreman} also indicates whether subgrid-scale models satisfy production property \hyperref[sec:modconstraintsSGSTKENicoud]{P2a} of Nicoud \etal~\cite{nicoudetal11} (\cf~\cref{sec:modconstraintsSGSTKENicoud}).
Indeed, one can
compare the size of the flow algebra of the different models with the number of two-component flows in Vreman's classification, as listed next to `2C flows'.
A more precise assessment of subgrid-scale models with respect to property \hyperref[sec:modconstraintsSGSTKENicoud]{P2a} can however be obtained by checking if 
their subgrid dissipation vanishes for all two-component flows,
as characterized by \cref{eq:2Cflowchar}.
This leads to the results provided in \cref{tab:modanalysis}.
With reference to property \hyperref[sec:modconstraintsSGSTKENicoud]{P2b}, we note that not many subgrid-scale models 
have been found
that vanish for states of pure axisymmetric strain.

In the context of reversibility of subgrid-scale models we remarked that most of the listed eddy viscosities are positive for all possible flow fields.
In fact, only $\nuesup{S3}$ can become negative, and only for certain values of the model parameter\replace{}{,} $p$.
Therefore, most of the discussed eddy viscosity models are time irreversible and consistent with the second law of thermodynamics~(\hyperref[eq:modreq2ndlaw]{P3}).
We note here that,
as a consequence,
backscatter cannot generally be captured by these models.
The gradient model, \cref{eq:modgrad}, can account for backscatter,
but it does so in a way that violates the second law of thermodynamics
and that causes simulations to blow up~\cite{vremanetal96,winckelmansetal01,berselliiliescu03}.

To 
test if Verstappen's minimum-dissipation condition for scale separation~(\hyperref[eq:modreqVerstappen]{P4}) is satisfied by
subgrid-scale models,
we will make a few simplifying assumptions.
First of all, 
for simplicity,
we will focus our attention on eddy viscosity models,
for which we will assume that eddy viscosities are constant over the filtering volume, $V_{ \testfilterlength }$.
Secondly, we will assume that the transport terms in \cref{eq:modreqVerstappen} can be neglected.
Any 
body forces
involving second derivatives of the velocity field are furthermore rewritten
in terms of first derivatives
using a Rayleigh quotient.
Finally, under the assumption of a finite viscosity, $\nu$, and with application of the midpoint rule to compute the volume integrals, \cref{eq:modreqVerstappen} reduces to
\begin{equation}
\label{eq:modreqVerstappeneddy}
\nue \geq - C \testfilterlength^2 \frac{ I_3 - I_4 }{ I_1 - I_2 }.
\end{equation}
Here, $C$ is a constant that relates to the Rayleigh quotient and that, 
similarly to the Poincaré constant, 
depends only on the shape of the filter volume, $V_{ \testfilterlength }$.
This constant is not known in general.
In the current work we have therefore determined whether the \emph{form} of eddy viscosity models is such that \cref{eq:modreqVerstappeneddy} can be satisfied. That is, can a value of a model's constant be found, such that \cref{eq:modreqVerstappeneddy} is satisfied.
This amounts to checking whether $\nue$ is finite whenever $-( I_3 - I_4)$ assumes a nonzero positive value.
The results of this exercise are shown in \cref{tab:modanalysis}.
It is to be noted that,
even for models that have the proper form,
\cref{eq:modreqVerstappeneddy} may not actually be satisfied
once practical values are used for the model constants.
Also note that \cref{eq:modreqVerstappeneddy} cannot be satisfied by subgrid-scale models that comply with property \hyperref[sec:modconstraintsSGSTKENicoud]{P2b}, because $-(I_3 - I_4 )$ does not vanish for the axisymmetric strain state.
Finally, we remark that the QR model only satisfies \cref{eq:modreqVerstappeneddy} if we assume that the filtering volume, $V_{ \testfilterlength }$, is a periodic box. 
This assumption, 
although employed in deriving the QR model~\cite{verstappenetal10,verstappen11,verstappenetal14},
was not made, here.

\subsection{Concluding remarks}
\label{sec:modelAnalysisConclusions}

The general view that we obtain from \cref{tab:modanalysis} and the above discussion is that 
the 
subgrid-scale models 
that have been considered 
so far
do not 
exhibit all the desired properties.
This can partly be understood from the fact that certain model constraints are not compatible with each other.
As remarked before in \cref{sec:modconstraintsSGSTKENicoud},
requirement \hyperref[sec:modconstraintsSGSTKENicoud]{P2a} of Nicoud \etal~cannot be satisfied simultaneously with Vreman's second model requirement~(\hyperref[eq:modreqVremanb]{P1b}).
Neither is production property \hyperref[sec:modconstraintsSGSTKENicoud]{P2b} compatible with Verstappen's minimum-dissipation condition (\hyperref[eq:modreqVerstappen]{P4}, also see \cref{eq:modreqVerstappeneddy} in \cref{sec:modanalysisSGSTKE}).
Furthermore, there seem to be some inherent limitations to velocity-gradient-based subgrid-scale models,
as scaling invariance~(\hyperref[eq:symmreq5]{S5}) cannot be satisfied without additional techniques like the dynamic procedure~\cite{oberlack97,razafindralandyetal07}.
Also 
no expression for the eddy viscosity was found that satisfies both of Vreman's model requirements~(\hyperref[eq:modreqVremana]{P1a}, \hyperref[eq:modreqVremanb]{P1b}),
but this may be due to the fact that we assumed the use of an isotropic filter in the computation of the turbulent stresses.
We believe, 
however,
that,
despite these observations,
there is room for improvement in the properties and, hence, the behavior, of subgrid-scale models that are based on the local velocity gradient.
In 
\cref{sec:modexamples}
we will give a few examples of such models.

We do note that
the incompatibilities between model constraints and the fact that even some very successful subgrid-scale models do not satisfy all the discussed requirements 
warrant an assessment of the practical importance and significance of each of the model requirements.
In this context we note that Fureby and Tabor~\cite{furebytabor97} performed an interesting study of the role of realizability in large-eddy simulations.

\section{Examples of new subgrid-scale models}
\label{sec:modexamples}

Having discussed the properties of some existing subgrid-scale models,
we now aim to illustrate how new
models for the turbulent stress tensor
can be constructed.
The model constraints of \cref{sec:modconstraints} will serve as our guideline in this process.
In this section we also show results of large-eddy simulations using a new eddy viscosity model.

\subsection{Derivation and properties}
As starting point for constructing new subgrid-scale models,
we 
take
the general class of models 
that are nonlinear in the local velocity gradient,
given by \cref{eq:modnonlin,eq:tensors,eq:modcoefficients,eq:tensorinvariants}.
Each of the model requirements of \cref{sec:modconstraints} can be used to restrict this class of models,
which leads to information about the functional dependence of the model coefficients, $\alpha_i$, \cref{eq:modcoefficients}, on the tensor invariants of \cref{eq:tensorinvariants}.
Here it is important to keep 
the limitations of velocity-gradient-based subgrid-scale models 
and 
the incompatibilities between model constraints, as discussed in \cref{sec:modelAnalysisConclusions}, in mind.

When compatible constraints are combined to restrict the general class of subgrid-scale models of \cref{eq:modnonlin}, the dependence of the model coefficients on the tensor invariants of \cref{eq:tensorinvariants} is 
not fully determined.
We thus obtain a class of 
subgrid-scale models.
The simplest models in this class
that exhibit the proper near-wall scaling behavior~(\hyperref[eq:modreqnearwallscal]{N})
have coefficients that depend only on the invariants of the rate-of-strain tensor, $I_1 = \tr{ \LESS^2 }$ and $I_3 = \tr{ \LESS^3 }$. 
For example (Nonlinear (NL) example model),
\begin{equation}
\label{eq:modNL}
\taumodsub{NL} = C_0 \filterlength^2 \frac{ I_3^4 }{ I_1^5 } I + C_1 \filterlength^2 \frac{ I_3^3 }{ I_1^4 } \LESS + C_4 \filterlength^2 \frac{ I_3^4 }{ I_1^6 } ( \LESS\LESOmega - \LESOmega\LESS ).
\end{equation}
Here, the $C_i$ are used to denote dimensionless model constants and, again, $\filterlength$ represents the subgrid characteristic (or filter) length.
With nondynamic constants, the above model satisfies all the symmetries of the Navier-Stokes equations, apart from scale invariance~(\hyperref[eq:symmreq5]{S5}).
Also time reversal invariance~(\hyperref[eq:symmreq1to3and7]{S7}) is satisfied, but, due to this, consistency with the second law of thermodynamics~(\hyperref[eq:modreq2ndlaw]{P3}) is lost.
The above subgrid-scale model is not realizable, mainly due to the appearance of the rate-of-rotation tensor in the third term on the right-hand side of \cref{eq:modNL} and the absence of this quantity in the model coefficient of the first term.
It seems reasonable to assume that
a more complex definition of the model coefficients could be found to solve this problem.

An interesting aspect of the above model is that it \replace{consists of}{contains} nonlinear terms that in general are not aligned with the rate-of-strain tensor. 
It can therefore describe nondissipative processes.
In fact,
the three terms of the above model are mutually orthogonal.
Therefore they each have their own physical significance.
The first term on the right-hand side of \cref{eq:modNL} models the generalized subgrid-scale kinetic energy, whereas the second term, the usual eddy viscosity term, describes dissipative processes.
The last term does not directly influence the subgrid dissipation 
and relates to energy transport among large scales of motion.
\replace{}{Given the number of model requirements for the production of subgrid-scale kinetic energy, \cf~\cref{sec:modconstraintsSGSTKE}, it is clear that subgrid-scale models are commonly characterized and assessed in terms of their dissipation properties.
As far as the authors are aware, it is far less common to characterize (let alone assess) subgrid-scale models in terms of transport of energy (do however take note of the work by Anderson and Domaradzki~\cite{andersondomaradzki12}).}
Nonlinear models of the form of \cref{eq:modNL} will\replace{}{, therefore,} be studied in subsequent work.

In view of the requirements of Nicoud \etal~(\hyperref[sec:modconstraintsSGSTKENicoud]{P2a}, \hyperref[sec:modconstraintsSGSTKENicoud]{P2b}), 
a possibly attractive quantity to base an eddy viscosity model on is the nonnegative quantity 
$4 ( I_5 - \frac{ 1 }{ 2 } I_1 I_2 ) = 4 ( \tr{ \LESS^2 \LESOmega^2 } - \frac{ 1 }{ 2 } \tr{ \LESS^2} \tr{ \LESOmega^2 } )$~\cite{silvisverstappen15b,silvisverstappen15c}.
This quantity
equals the (squared) magnitude of the vortex stretching,
$\LESS_{ij} \omega_j$~\cite{triasetal15},
where the components of the vorticity vector are related to the rate-of-rotation tensor via $\omega_i = -\epsilon_{ijk} \LESOmega_{jk}$ and $\epsilon_{ijk}$ again represents the Levi-Civita symbol.
The vortex stretching magnitude can serve as a correction factor for the dissipation behavior (damping the subgrid dissipation in locally laminar flows) and the near-wall scaling of the Smagorinsky model.
To that end, we first make the vortex-stretching magnitude dimensionless by dividing it by $-I_1 I_2$. 
This is a positive quantity that seems suitable because, like the vortex stretching magnitude, it depends quadratically on both the rate-of-strain and rate-of-rotation tensors.
Secondly, from the discussion of \cref{sec:modanalysisnearwallscal}, we deduce a power that ensures the proper near-wall scaling behavior.
This provides us with the following model, which we will refer to as the 
vortex-stretching-based 
(VS) eddy viscosity model,
\begin{equation}
\label{eq:modVS}
\taumoddevsub{ VS }
= -2 \nuesup{VS} \LESS = -2 ( \Csub{VS} \filterlength )^2 \sqrt{ 2 I_1 } \left( \frac{ I_5 - \frac{ 1 }{ 2 } I_1 I_2 }{ -I_1 I_2 } \right)^{3/2} \LESS.
\end{equation}
\replace{}{Defining $I_1 = \tr{ \LESS^2} = | \LESS |^2$ and $I_2 = \tr{ \LESOmega^2 } = - | \LESOmega |^2 = -\frac{ 1 }{ 2 } | \vec{\omega} |^2$, and employing matrix notation, we may also write
\begin{equation}
\label{eq:modVSvector}
\taumoddevsub{ VS }
= -2 ( \Csub{VS} \filterlength )^2 \frac{ 1 }{ 2 } | \LESS | \left( \frac{ | S \vec{\omega} | }{ | S | | \vec{\omega} | } \right)^3 \LESS.
\end{equation}
This shows that only the rate-of-strain tensor, $\LESS_{ij}$, and the vorticity vector, $\omega_i$, are required to compute the vortex-stretching-based eddy viscosity model.}

By construction the 
vortex-stretching-based
eddy viscosity model has the desired near-wall scaling behavior~(\hyperref[eq:modreqnearwallscal]{N}).
Furthermore, it vanishes only in two-component flows and in states of pure shear and pure rotation.
It has a positive value of eddy viscosity for all possible flow fields, so that
it is 
time irreversible
and 
consistent with the second law of thermodynamics~(\hyperref[eq:modreq2ndlaw]{P3}).
This model does not satisfy Verstappen's minimum-dissipation condition for scale separation~(\hyperref[eq:modreqVerstappen]{P4}), but it is to be noted that this is only due to one particular flow, the state of pure shear, for which $\nuesup{VS}$ vanishes and $-( I_3 - I_4)$ is finite (refer to \cref{eq:modreqVerstappeneddy}).

For comparison, the properties of the above nonlinear example model and the 
vortex-stretching-based
eddy viscosity model are
included in \cref{tab:modanalysis,tab:modconstraintsVreman}.

\subsection{Numerical tests}
\label{sec:numTests}
We will now test the vortex-stretching-based eddy viscosity model, \cref{eq:modVS}, in large-eddy simulations of decaying homogeneous isotropic turbulence and turbulent plane-channel flow.
These test cases and the numerical results that were obtained are described in detail below.
The simulations were performed using an incompressible Navier-Stokes solver that employs a symmetry-preserving finite-volume discretization on a staggered grid and a one-leg time integration scheme~\cite{verstappenveldman03}.
The results shown in what follows were obtained at second-order spatial accuracy.

\subsubsection{Decaying homogeneous isotropic turbulence}
\label{sec:numTestsHit}

We first consider 
large-eddy simulations 
of decaying homogeneous isotropic turbulence.
As reference for these simulations
we take the experimental data of Comte-Bellot and Corrsin (CBC)~\cite{comtebellotcorrsin71},
who
performed measurements of (roughly) isotropic turbulence
generated by a regular grid in a uniform air flow.
Of particular interest for our purposes are the energy spectra, that were measured at three different stations 
downstream of
the grid.

To link the experimental setup and the numerical simulation,
we imagine that
we are following the flow inside of a box that is moving away from the 
turbulence-generating grid
(with mesh size $M = 5.08 \units{cm}$)
at the mean velocity, $U_0 = 1000 \units{cm \: s^{-1}}$.
As such, the time in the numerical simulation plays the role of the distance from the grid in the experiment.
The length of the edges of the box, $\Lref = 11 M = 55.88 \units{cm}$, and a reference velocity $\uref = 27.19 \units{cm \: s^{-1}}$,
that corresponds to the kinetic energy content of the flow at the first measurement station, $\uref^2 = \frac{ 3 }{ 2 } \widebar{ u_1^2 }$,
are used to make quantities dimensionless~\cite{rozemaetal15}.
Taking the proper value of the viscosity,
$\nu = 0.15 \units{cm^2 \: s^{-1}}$,
we thus obtain a Reynolds number of $\ReNr = 10,129$.

To allow for a comparison between numerical results and the experimental data,
we 
further
have to ensure a proper initial condition is used in the simulations.
To that end, we 
follow the procedure outlined by Rozema \etal~\cite{rozemaetal15},
and use
the MATLAB scripts that these authors 
were so kind to make available.%
\footnote{ See \url{http://web.stanford.edu/~hjbae/CBC} for a set of MATLAB scripts that can be used to generate initial conditions for large-eddy simulations of homogeneous isotropic turbulence. }
In this procedure, 
first an incompressible velocity field with random phases is created that fits the energy spectrum that was measured at the first station in the CBC experiment~\cite{kwaketal75}.
This velocity field is then fed into a preliminary large-eddy simulation with the QR model, to adjust the phases.
After 
a rescaling operation~\cite{kangetal03} a velocity field is obtained that has the same energy spectrum 
as the flow in the first measurement station.
This velocity field 
is suitable to be used
as initial condition
in a large-eddy simulation.
Energy spectra from the simulation can now be compared to the experimental spectra obtained in measurement stations two and three.

We performed large-eddy simulations of decaying homogeneous isotropic turbulence using the 
vortex-stretching-based
eddy viscosity model of \cref{eq:modVS}
on a uniform $64^3$ Cartesian computational grid with periodic boundary conditions.
The value of the model constant, $\Csub{VS}$, was estimated by requiring that the average dissipation due to the model,
$\avg{ \Dmodsub{VS} } = 2 \avg{ \nuesup{VS} I_1 }$,
matches the average dissipation of the Smagorinsky model,
$\avg{ \Dmodsub{S} } = 2 \avg{ \nuesup{S} I_1 }$\replace{.}{~\cite{nicoudducros99,nicoudetal11,triasetal15}.
More specifically,
\begin{equation}
\Csub{VS}^2 \approx \Csub{S}^2 \avg{ \sqrt{ 2 I_1 } I_1 } / \avg{ \sqrt{ 2 I_1 } \left( \frac{ I_5 - \frac{ 1 }{ 2 } I_1 I_2 }{ -I_1 I_2 } \right)^{3/2} I_1 }.
\end{equation}
Here,}
$\avg{ \cdot }$ indicates an \replace{}{ensemble }average over a large number of \replace{random }{}velocity gradients.\replace{\cite{nicoudetal11,triasetal15}}{ These can, for example, come from simulations of homogeneous isotropic turbulence~\cite{nicoudducros99}.
In the current work, a large number of `synthetic' velocity gradients was used, given by traceless random $3 \times 3$ matrices~\cite{nicoudetal11,triasetal15} sampled from a uniform distribution.}
A MATLAB script that performs this estimation of the constants of eddy viscosity models has been made freely available.%
\footnote{ See \url{https://bitbucket.org/mauritssilvis/lestools} for a set of MATLAB scripts that can be used to estimate the model constants of eddy viscosity models for large-eddy simulation. }
In this case it provides $\Csub{VS} \approx 3.4 \Csub{S} \approx 0.58$, for a Smagorinsky constant of $\Csub{S} = 0.17$.
Subsequently, the model constant was fine-tuned in such a way that the energy spectra from the simulations and the experiment show the best comparison for wavenumbers just above the numerical cutoff.
This led to $\Csub{VS} \approx 0.68$.

\begin{figure}
    \centering
    \includegraphics[scale=1,bb=0pt 0pt 473pt 184pt,viewport=0pt 0pt 473pt 184pt]{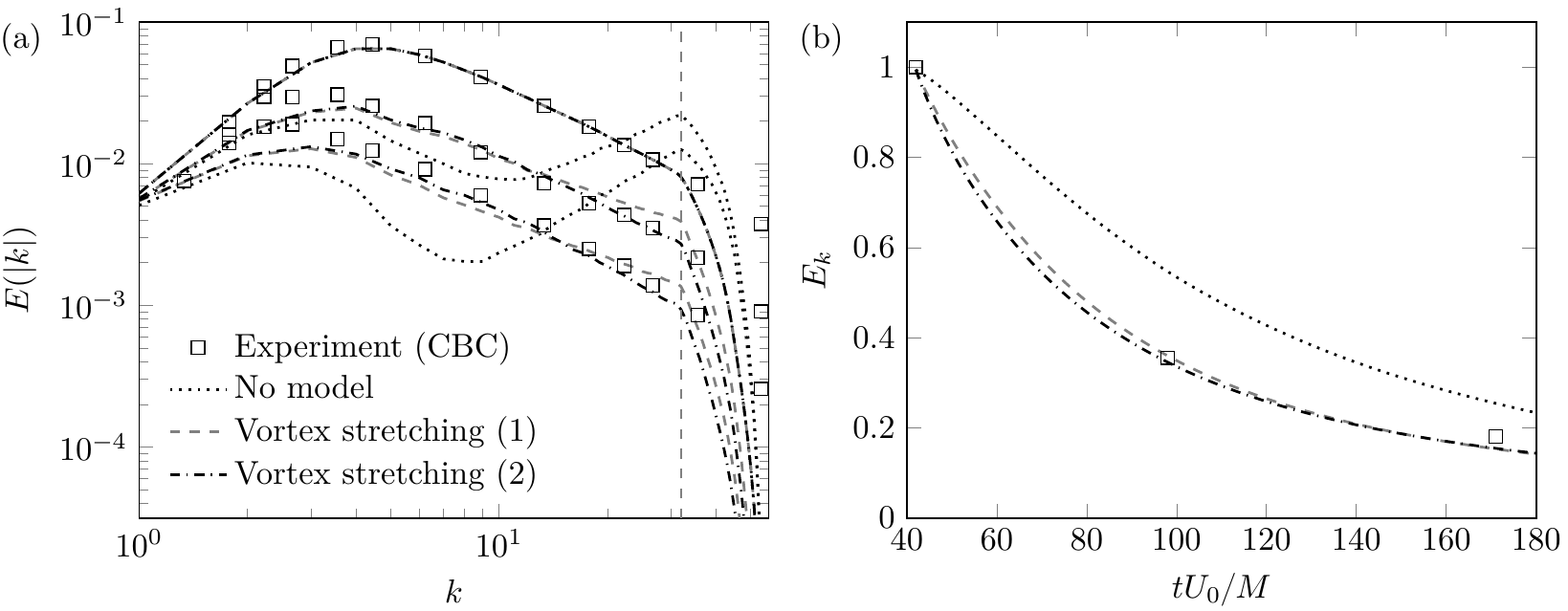}
    \caption{
        \label{fig:HIT_Ek_Ekin}
        (a) Three-dimensional kinetic energy spectra as a function of computational wavenumber at the three measurement stations and (b) decay of the total normalized resolved kinetic energy, as obtained from large-eddy simulations of homogeneous isotropic turbulence on a $64^3$ grid. Results are shown for simulations without a subgrid-scale model (dotted line) and with the vortex-stretching-based eddy viscosity model, Eq.~(\ref{eq:modVS}), with $C_{ \mathrm{ VS } } \approx 0.58$ (1) (dotted line) and $C_{ \mathrm{ VS } } \approx 0.68$ (2) (dash-dotted line). The experimental data from the experiment by Comte-Bellot and Corrsin (CBC)~\cite{comtebellotcorrsin71} are shown for reference (squares).
        }
\end{figure}

\cref{fig:HIT_Ek_Ekin} shows the resulting energy spectra and the normalized total resolved kinetic energy. 
For comparison, the experimental data from the CBC experiment and results from large-eddy simulations without a model are provided.
The resolved kinetic energy of the CBC experiment was estimated from discrete $64^3$ velocity fields with random phases that fit the three experimental energy spectra.
A good match between results from the large-eddy simulation and experimental data is obtained.
\replace{}{In fact, the vortex-stretching-based eddy viscosity model performs as least as good in simulations of homogeneous isotropic turbulence as Vreman's model, \cref{eq:modVreman}.
The currently obtained results, with $\Csub{VS} \approx 0.68$, are practically indistinguishable from the energy spectra and decay of kinetic energy that are predicted by simulations using Vreman's model with $\Csub{V} \approx 0.27$ (not shown).
\cref{fig:HIT_Ek_Ekin} also provides an indication of the sensitivity of simulation results with respect to the model constant, $\Csub{VS}$.
A change in the model constant of more than 10\%, from $\Csub{VS} \approx 0.58$ to $\Csub{VS} \approx 0.68$, still leads to a reasonable prediction of the three-dimensional kinetic energy spectra and a very satisfying prediction of the decay of kinetic energy.}

\subsubsection{Plane-channel flow}
\label{sec:numTestsCf}

Next, we focus on large-eddy simulations of a turbulent plane-channel flow.
The reference data for this test case come from the 
\replace{direct numerical simulation}{Direct Numerical Simulation} (DNS) performed by \replace{Moser, Kim and Mansour.}{Moser \etal}~\cite{moseretal99}.
Among other statistical quantities, these authors collected the mean velocity, $\avg{ u_i }$, and the Reynolds stresses, $R_{ij} = \avg{ u_i u_j } - \avg{ u_i } \avg{ u_j }$, where $\avg{ \cdot }$ now indicates an average over time and the two homogeneous directions.

The large-eddy simulations were performed using the vortex-stretching-based eddy viscosity model, \cref{eq:modVS}\replace{}{, and using Vreman's model, \cref{eq:modVreman}}.
To match the reference data, a constant pressure gradient is prescribed to maintain a friction Reynolds number\replace{,}{} based on the wall shear stress\replace{,}{} of $\Retau \approx 590$.
\replace{Furthermore, t}{T}he domain size is taken to be $(L_1, L_2, L_3) = (2 \pi \channelHalfWidth, 2 \channelHalfWidth, \pi \channelHalfWidth)$, where $\channelHalfWidth$ represents the channel half-width.
Again a $64^3$ grid is used, but
here it is only taken to be uniform and periodic in the streamwise ($x_1$) and spanwise ($x_3$) directions,
whereas the grid is stretched in the wall-normal direction ($x_2$).
Because of the stretching and anisotropy of the grid, 
characterized by $\Delta x_1^+ \approx 58$, \replace{$\Delta x_2^+ \approx 4 - 61$}{$\Delta x_2^+ \approx 2 - 61$} and $\Delta x_3^+ \approx 29$, 
the subgrid characteristic length scale was defined according to Deardorff's cube root of the \replace{cell volume, $\delta = ( \Delta x_1 \Delta x_2 \Delta x_3 )^{1/3}$\cite{deardorff70}.}{cell volume, $\delta = ( \Delta x_1 \Delta x_2 \Delta x_3 )^{1/3}$~\cite{deardorff70}.}

\begin{figure}
    \centering
    \includegraphics[scale=1,bb=0pt 0pt 467pt 367pt,viewport=0pt 0pt 467pt 367pt]{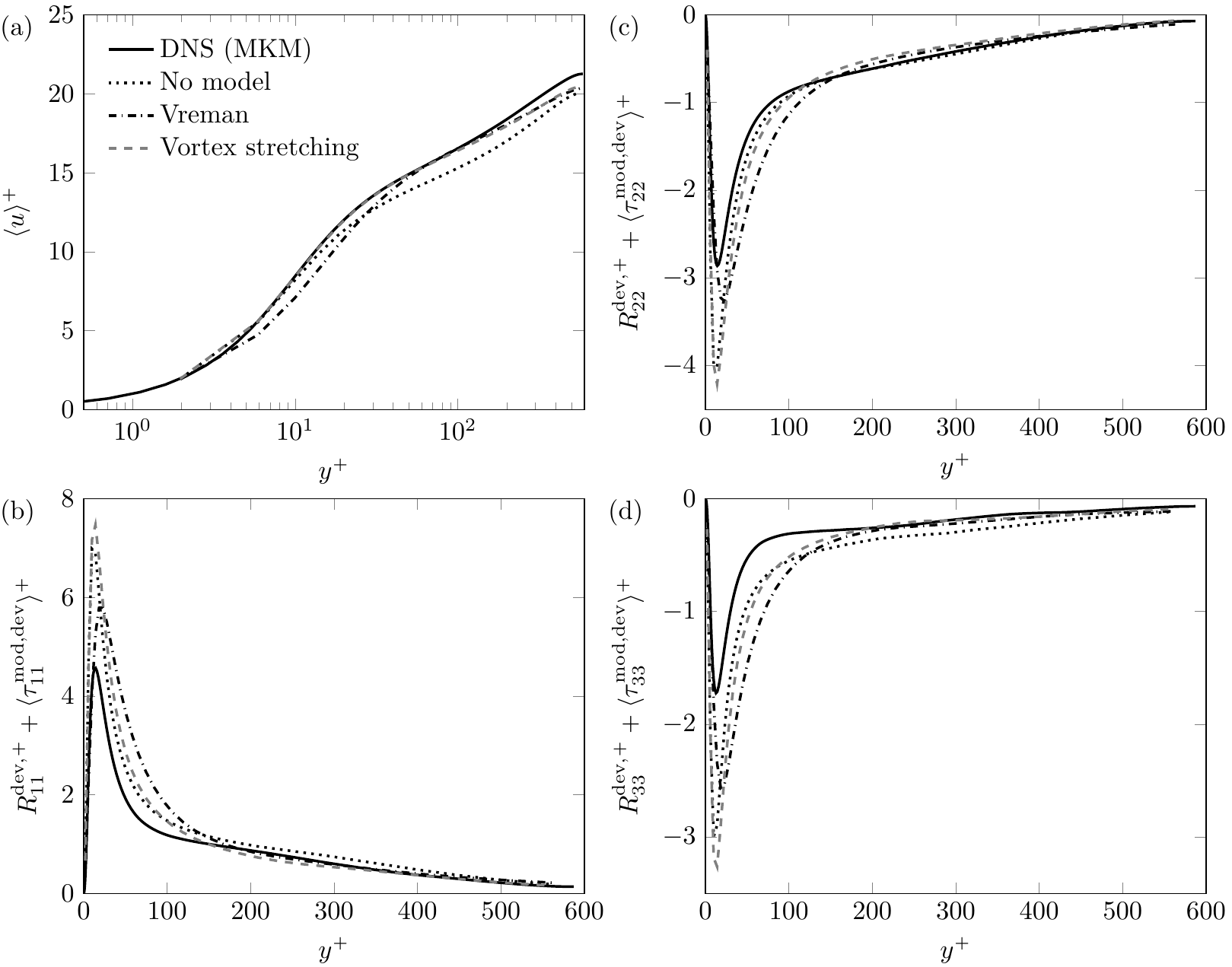}
    \caption{
        \label{fig:CF_u_Rii}
        (a) Mean velocity profile and (b) streamwise, (c) wall-normal and (d) spanwise deviatoric Reynolds stresses compensated by the average model contribution, as obtained from large-eddy simulations of plane-channel flow at $Re_{\tau} \approx 590$ on a $64^3$ grid. Results are shown for simulations without a subgrid-scale model (dotted line), with Vreman's model, Eq.~(\ref{eq:modVreman}), with $C_{ \mathrm{ V } } \approx 0.28$ (dash-dotted line) and with the vortex-stretching-based eddy viscosity model, Eq.~(\ref{eq:modVS}), with $C_{ \mathrm{ VS } } \approx 0.55$ (dashed line). Results from Direct Numerical Simulation (DNS)\cite{moseretal99} are shown for reference (solid line). All results are shown in wall units.
        }
\end{figure}

\Cref{fig:CF_u_Rii} shows the mean velocity profile and the Reynolds stresses as obtained from the large-eddy simulation\replace{}{s}.
The DNS data from \replace{Moser, Kim and Mansour}{Moser \etal}~\cite{moseretal99} and a no-model large-eddy simulation are shown for reference.
To allow for a fair comparison between the Reynolds stresses from the direct numerical simulation and \replace{the large-eddy simulation, using the traceless vortex-stretching-based eddy viscosity model}{from large-eddy simulations using traceless eddy viscosity models such as the vortex-stretching-based eddy viscosity model and Vreman's model}, we show only the deviatoric Reynolds stresses and compensate for the average model contribution~\cite{winckelmansetal02}.
All quantities are expressed in wall units, based on the friction velocity, $\utau$, and the channel half-width, $\channelHalfWidth$.

For the vortex-stretching-based eddy viscosity model\replace{}{,} results are shown with \replace{$\Csub{VS} \approx 0.58$, }{$\Csub{VS} \approx 0.55$.
For this value of the model constant, which is remarkably close to $\Csub{VS} \approx 0.58$ (}the value that was obtained from matching the average model dissipation with that of the Smagorinsky model (see \cref{sec:numTestsHit})\replace{.
For this value of the model constant, }{), }the mean velocity in the near-wall region is predicted \replace{remarkably}{very} well.
\replace{}{For Vreman's model, results are shown with $\Csub{V} \approx 0.28$, which is the value of the model constant for which the bulk velocity (the spatial average of the mean velocity) predicted by both subgrid-scale models is approximately equal.
Vreman's model fails to capture the behavior of the mean velocity near the wall.}
The mean velocity in the center of the channel is underpredicted\replace{}{ by both models}, but it is noted here that it seems to be a common deficiency of eddy viscosity models to not be able to predict the inflection of the mean velocity starting around $y^+ \approx 100$.
\replace{}{Note that both models are relatively robust when it comes to changes in the model constant. Changing $\Csub{VS}$ or $\Csub{V}$ by around 10\% only causes a 1\% change in the bulk velocity.
For comparison, the simulations without a model had a 4\% smaller bulk velocity than the simulations with a subgrid-scale model.}

\replace{In the Reynolds stress profiles we observe that the streamwise stresses are overpredicted}{The Reynolds stress profiles show that both subgrid-scale models overpredict the streamwise stresses}, while the other diagonal \replace{components}{stresses} are underpredicted.
Again this seems to be a common feature of eddy viscosity models.
\replace{The location of the near-wall peak in the stresses is predicted well.}{Note, however, that Vreman's model predicts Reynolds stresses with rather broad near-wall peaks, located too far from the wall, while the vortex-stretching-based eddy viscosity model predicts well the location and width of the near-wall peaks.
The near-wall scaling behavior of both subgrid-scale models may explain these differences.}

\begin{figure}
    \centering
    \includegraphics[scale=1,bb=0pt 0pt 233pt 185pt,viewport=0pt 0pt 233pt 185pt]{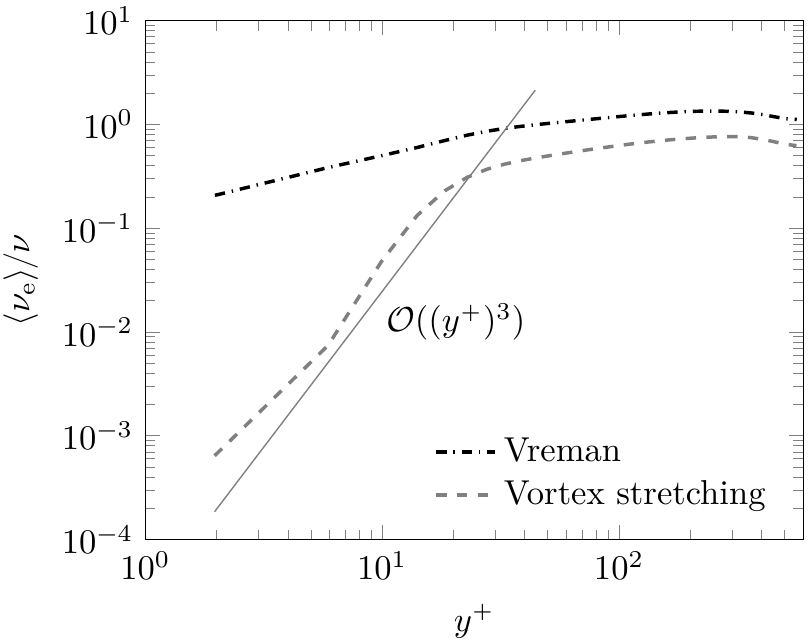}
    \caption{
        \label{fig:CF_nue}
        \replace{}{Average eddy viscosity in units of the kinematic viscosity, $\nu$, as obtained from large-eddy simulations of plane-channel flow at $\Retau \approx 590$ on a $64^3$ grid. 
        Results are shown for simulations using the vortex-stretching-based eddy viscosity model, \cref{eq:modVS}, with $\Csub{VS} \approx 0.55$ (dashed line) and Vreman's model, \cref{eq:modVreman}, with $\Csub{V} \approx 0.28$ (dash-dotted line).
        The desired near-wall scaling behavior of the eddy viscosity is indicated (solid line).}
    }
\end{figure}

\replace{}{The near-wall scaling of the vortex-stretching-based eddy viscosity model and of Vreman's model is studied in \cref{fig:CF_nue}, which shows the average eddy viscosity measured during the large-eddy simulations.
The vortex-stretching-based eddy viscosity model exhibits the desired near-wall scaling, $\nue = \mathcal{O}( (y^+)^3 )$, see \cref{sec:modanalysisnearwallscal}, which ensures that the subgrid dissipation is not too large close to the wall.
On the other hand, Vreman's model shows a different near-wall scaling behavior, close to  $\nue = \mathcal{O}( y^+ )$, allowing for much higher values of eddy viscosity and, thus, a larger subgrid dissipation close to the wall.
This may explain the problems Vreman's model has with predicting the mean velocity and Reynolds stresses close to the wall.
Improvements in the near-wall behavior of Vreman's model may, however, also be obtained by using the original anisotropic implementation of the model~\cite{vreman04}, rather than the implementation of \cref{eq:modVreman}, which is based on an isotropic eddy viscosity with an anisotropic subgrid characteristic length scale, in this case given by Deardorff's length scale~\cite{deardorff70}.
}

\replace{All in all, it is rather remarkable how well the vortex-stretching-based eddy viscosity model works in combination with Deardorff's length scale and a simple estimation procedure for the model constant based on the average model dissipation.}{In conclusion, we have seen how the model constraints of \cref{sec:modconstraints} can be used to derive new subgrid-scale models.
An example of such a model, the vortex-stretching-based eddy viscosity model, \cref{eq:modVS}, was tested in large-eddy simulations of homogeneous isotropic turbulence and a turbulent plane-channel flow.
Good predictions of three-dimensional kinetic energy spectra and the decay of kinetic energy were obtained for the former test case, as well as for the mean velocity and Reynolds stresses in the latter test case.
Although improvements are possible, in the form of a better prediction of the mean velocity in the center of the channel and the height of the near-wall peaks in the Reynolds stresses, we believe we have obtained encouraging results, that show the power of the framework of model constraints discussed in this work to design new subgrid-scale models with built-in desirable properties.}

\section{Summary}
\label{sec:concldisc}

We studied the construction of subgrid-scale models for large-eddy simulation of incompressible turbulent flows.
In particular, we aimed to consolidate a systematic approach of constructing subgrid-scale models.
This approach is based on the idea that it is desirable that subgrid-scale models are consistent with important mathematical and physical properties of the Navier-Stokes equations and the turbulent stresses.
We first discussed in detail several of these properties, namely: the symmetries of the Navier-Stokes equations, and the near-wall scaling behavior, realizability and dissipation properties of the turbulent stresses.
In the last category we focused on Vreman's study of the dissipation behavior of the turbulent stresses~\cite{vreman04}, the physical model requirements of Nicoud \etal~\cite{nicoudetal11}, consistency with the second law of thermodynamics and Verstappen's dissipation considerations relating to scale separation in large-eddy simulation~\cite{verstappen11,verstappen16}.
We furthermore outlined the requirements that subgrid-scale models have to satisfy in order to preserve these mathematical and physical properties.

As such, a framework of model constraints arose, that we subsequently applied to investigate the properties of existing subgrid-scale models.
We focused specifically on the analysis of the behavior of subgrid-scale models that depend locally on the velocity gradient.
These models satisfy some desired properties by construction, such as Galilean invariance and rotational invariance.
Also, most subgrid-scale models of this form are consistent with the second law of thermodynamics. 
Other properties are only included in some of the existing models.
For example, the WALE model~\cite{nicoudducros99}, the $\sigma$ model~\cite{nicoudetal11} and the S3PQR models~\cite{triasetal15} have the proper near-wall scaling behavior, whereas the other models under study do not.
The recently developed anisotropic minimum-dissipation (AMD) model~\cite{rozemaetal15} was designed to exhibit a particular dissipation behavior that leads to scale separation between large and small scales of motion, a property that is shared by some, but not all other models.

Thus, the subgrid-scale models that we considered do not generally satisfy all the desired properties.
This can partly be understood from the fact that some model constraints, particularly dissipation properties, are not compatible with each other.
We mentioned that the physical requirement of Nicoud \etal~\cite{nicoudetal11}, that a model's subgrid dissipation should vanish for all two-component flows, does not match with Vreman's requirements derived from the mathematical properties of the turbulent stress tensor~\cite{vreman04}.
Also, Verstappen's requirement for scale separation~\cite{verstappen11} can only be satisfied if a model has a nonzero subgrid dissipation for the pure axisymmetric strain, contrary to another requirement of Nicoud \etal~\cite{nicoudetal11}.
We furthermore remarked that there seem to be some inherent limitations to subgrid-scale models that are based on the velocity gradient, as scaling invariance cannot be satisfied in the current formulation, where a length scale based on the local grid size is employed.
Moreover, no velocity-gradient-based expression for the eddy viscosity was found that satisfies both of Vreman's model requirements.
This, however, may be due to our assumption of a filter that conforms to the symmetries of the Navier-Stokes equations (an isotropic filter) in computing the dissipation behavior of the true turbulent stresses.
Despite these observations, we believe that there is room for improvement in the properties and, hence, the behavior of subgrid-scale models derived from the velocity gradient.
The current work provides several suggestions in this way.

We showed how compatible model constraints can be combined to derive new subgrid-scale models that have desirable properties built into them.
First, we illustrated how to construct a subgrid-scale model that is nonlinear in the velocity gradient, based on the invariants of the rate-of-strain tensor.
Then we proposed a new eddy viscosity model based on the vortex stretching magnitude, to correct for the near-wall scaling and dissipation behavior of the Smagorinsky model.
This new model has several interesting properties: it has the desired near-wall scaling behavior and it vanishes only in two-component flows, and in states of pure shear and pure rotation.
This vortex-stretching-based eddy viscosity model was tested successfully in large-eddy simulations of decaying homogeneous isotropic turbulence and turbulent plane-channel flow.

With the current work we hope to have consolidated systematic approaches for the assessment of existing and the creation of new subgrid-scale models for large-eddy simulation. 
In future work, we would like to investigate in more detail constraints for the construction of subgrid-scale models.
For example, incompatibilities between model requirements and the observation that even some very successful subgrid-scale models do not satisfy all the discussed requirements, warrant an assessment of the practical importance and significance of each of the model requirements.
Furthermore, the fact that the velocity-gradient-based subgrid-scale models we considered here do not comply with scale invariance, calls for a detailed analysis of the symmetry preservation properties of the dynamic procedure~\cite{germanoetal91}, of other techniques that provide estimates of the turbulent kinetic energy and energy dissipation rate like the integral-length scale approximation~\cite{piomellietal15,rouhietal16}, and of flow-dependent definitions of the subgrid characteristic length scale in general.
We would also like to investigate the behavior of subgrid-scale models that contain terms that are nonlinear in the rate-of-strain and rate-of-rotation tensors, particularly to \replace{determine if they can}{study their ability to} describe transport processes in flows.
Finally, we will focus on devising new constraints for the construction of subgrid-scale models.

\paragraph{Acknowledgments} The authors thankfully acknowledge Professor Martin Oberlack for stimulating discussions during several stages of this project.
Professor Michel Deville is kindly acknowledged for sharing his insights relating to nonlinear subgrid-scale models and realizability.
Theodore Drivas and Perry Johnson are thankfully acknowledged for their valuable comments and criticisms on \replace{a preliminary version}{preliminary versions} of this paper. 
Portions of this research have been presented at the 15th European Turbulence Conference, August 25-28th, 2015, Delft, The Netherlands, and at the 4th International Conference on Turbulence and Interactions, November 2nd-6th, 2015, Cargèse, Corsica, France.
This work is part of the Free Competition in Physical Sciences, which is financed by the Netherlands Organisation for Scientific Research (NWO). 
M.H.S. gratefully acknowledges support from the Institute for Pure and Applied Mathematics (Los Angeles) for visits to the ``Mathematics of Turbulence'' program during the fall of 2014.